\documentclass[9pt]{elife}

\usepackage{lipsum} %
\usepackage[version=4]{mhchem}
\usepackage{siunitx}
\usepackage{soul}
\usepackage{amsmath}
\DeclareMathOperator\erf{erf}
\DeclareSIUnit\Molar{M}

\title{Continuous odor profile monitoring to study olfactory navigation in small animals}

\author[1\authfn{1}]{Kevin S. Chen}
\author[2\authfn{1}]{Rui Wu}
\author[2,3,*]{Marc H. Gershow}
\author[1,4,*]{Andrew M. Leifer}
\affil[1]{Princeton Neuroscience Institute, Princeton University, New Jersey, United States}
\affil[2]{Department of Physics, New York University, New York, United States}
\affil[3]{Center for Neural Science, New York University, New York, United States}
\affil[4]{Department of Physics, Princeton University, New Jersey, United States}

\corr{mhg4@nyu.edu}{MHG}
\corr{leifer@princeton.edu}{AML}

\contrib[\authfn{1}]{These authors contributed equally to this work}

\begin{document}

\maketitle

\begin{abstract}

Olfactory navigation is observed across species and plays a crucial role in locating resources for survival. In the laboratory, understanding the behavioral strategies and neural circuits underlying odor-taxis requires a detailed understanding of the animal's sensory environment. For small model organisms like \textit{C. elegans} and larval \textit{D. melanogaster}, controlling and measuring the odor environment experienced by the animal can be challenging,  especially for airborne odors, which are subject to subtle effects from airflow, temperature variation, and from the odor's adhesion, adsorption or reemission. 
Here we present a method to flexibly control and precisely measure airborne odor concentration in an arena with agar while imaging animal behavior. Crucially and unlike previous methods, our method allows continuous monitoring of the odor profile during behavior. We construct stationary chemical landscapes in an odor flow chamber through spatially patterned odorized air. The odor concentration is measured with a spatially distributed array of digital gas sensors. Careful placement of the sensors allows the odor concentration across the arena to be accurately inferred and continuously monitored at all points in time. We use this approach to measure the precise odor concentration that each animal experiences as it undergoes chemotaxis behavior and report chemotaxis strategies for \textit{C. elegans} and \textit{D. melanogaster} larvae populations under different spatial odor landscapes.

\end{abstract}

\section{Introduction}
 
Odor-guided navigation is common across the animal kingdom \citep{Baker2018-ys}. Olfactory cues inform an animal of its location in a natural environment \citep{Boie2018-eb}, and allow it to adjust its locomotion to navigate an odor landscape in a goal directed manner \citep{bargmann1991chemosensory,berg1972chemotaxis,aceves1979learning}. Odor guided navigation is an ethologically relevant task that is important for the animal's survival, and it has been a useful framework with which to study genes and circuits underlying sensory-motor transformations \citep{calhoun2017quantifying, clark2013mapping}.
Small model organisms navigate continuous gradients established by the spread of odorants from their sources due to diffusion and drift. 
How animals interpret these gradients and use them to inform their actions remains an active and productive area of research especially in genetic model systems like \textit{C. elegans} and \textit{Drosophila melanogaster} \citep{bargmann1991chemosensory,aceves1979learning,Levy2020-oh,mattingly2021escherichia, Gomez-Marin2011-ok,Gepner2015-wm}

A major challenge is to quantitatively relate the animal's behavior to the precise olfactory cue that the animal experiences moment-by-moment.
Therefore it is critical to precisely control the odor environment and record the sensory cues experienced by these animals. The need to \emph{control} and \emph{measure} odorants still pose a formidable challenge. While many techniques exist to either present or measure odors in a lab environment, no technique currently exists for precise control and continuous monitoring of an odor landscape. 

All approaches to generate odor landscapes in a lab environment must contend with the odor's diffusivity and interaction with other substrates. Early approaches to \textit{control} odor concentration relied on passive diffusion to construct a quasi-stationary spatial odor gradient, for example by adding a droplet of odorant in a petri dish in a ``droplet assay'' \citep{Louis2008-ju,Iino2009-al,Pierce-Shimomura1999-nt, monte1989characterization}. Diffusion places severe limits on the space of possible landscapes that can be created and on the timescales over which they are stable, and the created odor profile is sensitive to adsorption of odor to surfaces, absorption into the substrate, temperature gradients, and air currents, all parameters that are difficult to measure, model, or control. Microfulidics allow water-soluble odors to be continuously delivered to a chamber in order to provide spatiotemporal control \citep{chronis2007microfluidics, Albrecht2011-fj, lockery2008artificial}. Microfluidics devices, however, are limited in extent, require water-soluble odors and must be tailored-designed to the specific attributes of the animal's size and locomotion. While a post array has been shown to support \textit{C. elegans} locomotion, no microfluidic device has been demonstrated to support olfactory navigation of \textit{Drosophila} larvae, for example. 

We previously reported a macroscopic gas-based active flow cell that uses parallel flow paths to construct temporally stable odor profiles \citep{Gershow2012-nt}. That approach allows for finer spatiotemporal control of the odor gradient, is compatible with \textit{Drosophila} larvae, and works with volatile airborne-based odor cues. This device used an array of solenoid valves to generate programmable odor profiles, but perhaps because of its complexity has not been widely adopted.

Most methods to create an odor landscape do not provide a means for knowing or specifying the spatiotemporal odor concentration. In other words while an experimenter may know that some regions of an area have higher odor concentrations, they cannot quantify the animal's behavior given a precise concentration of the odor. This limits the ability to quantitatively characterize sensorimotor processing. To address this shortcoming, various methods have been proposed to \textit{measure} odor concentration across space. For example, gas samples at specific locations could be taken and measured offline \citep{Yamazoe-Umemoto2018-nx}. In one of the most comprehensive measurements to date, Louis and colleagues
\citep{Louis2008-ju, tadres2022depolarization} used infra-red spectroscopy to measure the spatial profile of a droplet based odor gradient. 

In all of these cases, measurements were performed offline, not during animal behavior, and the odor concentration was assumed to be the same across repeats of the same experiment, and when animals are present. But even a nominally stable odor landscape is subject to subtle but significant disruptions over time from small changes in airflow, from temperature variation, and from the odor's interaction with the substrate, which can include absorption, adhesion, and reemission \citep{Gorur-Shandilya2019-me, Yamazoe-Umemoto2018-nx,tanimoto2017calcium,Yamazoe-Umemoto2015-ru}.
This is challenging to account for and control within a single behavior experiment, and is even more difficult to account for across multiple instances of such experiments. Additional variability also arises across experiments as a result of the introduction of animals, changes to agar substrates, and alteration in humidity or other environmental conditions. To recover the odor concentration that an animal experienced, there is a need to measure odor concentration and animals' behaviors concurrently.

Our previously reported flow cell used a photo-ionization detector (PID) sensor moved across the lid before behavioral experiments to measure the odor concentration across space at a single point in time \citep{Gershow2012-nt}. During experiments, the total concentration of odor in the chamber was monitored concurrently with measurements of behavior. While this provided some assurances that the overall odor concentration was relatively stable, it did not provide any spatial information concurrently with behavior measurements.

Here we present a new flow chamber and a new multi-sensor odor array that addresses these prior limitations and can be used for measurement of the odor gradient with high spatial and temporal resolution. The array of sensors can be used two ways: the full array can be used to measure the generated gradient throughout the extent of the chamber, or parts of the array can be used on the borders to monitor, \textit{during behavioral experiments}, the odor profile in the chamber. By varying flow rates and the sites of odor introduction, we show a variety of odor profiles can be generated and stabilized.

To demonstrate the utility of the apparatus, we applied this instrument to quantitatively characterize the sensorimotor transformation underlying navigational strategies used by \textit{C. elegans} and \textit{D. melanogaster} larva to climb up a butanone odor gradient. Butanone is a water-soluble odorant found naturally in food sources \citep{worthy2018identification} that is often used in odor-guided navigation studies \citep{bargmann1993odorant, Levy2020-oh, Cho2016-is, Torayama2007-qi}. We show that the agar gel used during behavioral experiments greatly disrupts an applied butanone gradient, and we demonstrate a pre-equilibration protocol allowing generation of stable gradients taking into consideration the effects of agar. Moreover we monitor these gradients during ongoing behavior measurements via continuous measurements of the odor profile along the boundaries of the arena. %

Using these stable and continuously measured butanone gradients, we measure odor-guided navigation in animals by tracking their posture and locomotion as they navigate the odor landscape. We record chemotaxis behavior and identify navigation strategies in response to the changing  odor concentration they experience. 
In \textit{C. elegans}, we observe the presence of navigational strategies that were reported in other sensory-guided navigation conditions, such as salt chemotaxis \citep{Iino2009-al, Dahlberg2020-ip, Luo2014-pc}. These two strategies are: a biased random turn, known as a pirouette \citep{Pierce-Shimomura1999-nt}, and a gradual veering, known as weathervaning \citep{Iino2009-al, Izquierdo2015-la}. %
In \textit{Drosophila melanogaster} larvae, we identify runs followed by directed turns \citep{Gershow2012-nt, Louis2008-ju, Gomez-Marin2011-ok}. %
By using concurrent measurements of behavior and odor gradient we characterize olfactory navigation in these small animals on agar with known butanone odor concentrations, which for \textit{C. elegans} has not been reported before.

\section{Results}

We developed new methods both for generating and measuring odor gradients which we describe here. The systems are modular, scalable, and flexible. The components, which can be used independently of each other, can be fabricated directly from provided files using online machining services, or the provided plans can be modified for other geometries. 

\subsection{Flow chamber for generating spatiotemporal patterns of airborne odors}

We first sought to develop a method of creating odor gradients that satisfied the following criteria:
\begin{enumerate}
    \item The spatial odor profile should be \textit{controllable}. Varying control parameters (e.g. flow rates, tubing connections) should result in predictable changes to the resulting odor landscape. 
    \item The odor profile should be \textit{stable} and \textit{verifiable}. The same spatial profile should be maintained over the course of an experiment lasting up to an hour, and this should be verifiable via concurrent measurements during behavior experiments.
    \item The apparatus should be \textit{straightforward} to construct and to use, and \textit{flexible} to adapt to various experimental configurations, including using with either \textit{C. elegans} or \textit{Drosophila} larva, and with agar arenas of various sizes.
\end{enumerate}

\begin{figure}
\begin{fullwidth}
\includegraphics[width=0.85\linewidth]{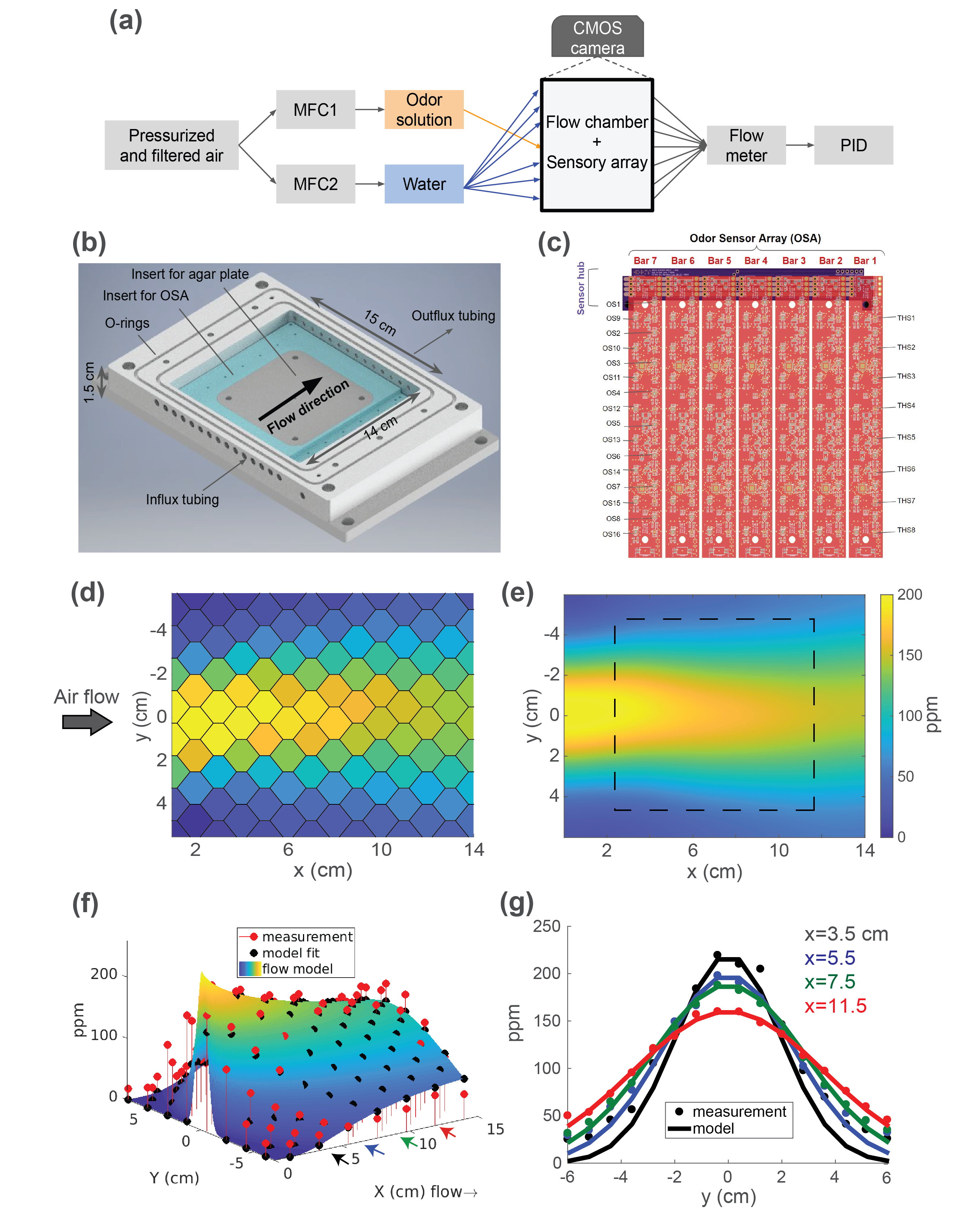}
\caption{\textbf{Odor flow chamber with controlled and measured odor concentration.} \textbf{(a)} Schematic of airflow paths. Airflow paths for odor solution and water are controlled separately by mass flow controllers (MFCs) and spatially arranged into the odor chamber. The outflux from the chamber connects to a flow meter and photo-ionization detector (PID). \textbf{(b)} Flow chamber design. \textbf{(c)} Odor sensory array (OSA). Seven odor sensor bars are connected to a sensor hub. Each bar has 16 odor sensors (OS) and 8 temperature/humidity sensors (THS). Measured odor concentrations from the OSA of a spatially patterned butanone odor concentration shown in \textbf{(d)} for each sensor, \textbf{(e)} interpolated across the arena with square dashed line indicating the area where agar and animals are placed. \textbf{(f)} A two-parameter analytic flow model fit to measurement. \textbf{(g)} Cross-sections from (f) at at 4 different x-axis positions (show as colored arrows). Sensor readouts are overlaid points on the smooth curves from a 1D diffusion model. \label{fig:fig1}
}
\figsupp[Long-duration calibration confirms stable control and measurements.]{Long-duration calibration confirms stable control and measurements. (a, top) Odor flow rate driven by the MFC, (a, middle) raw reading of an odor sensor (OS) located in the chamber, (a, bottom) odor concentration readout from the photo ionization detector (PID) located at the outlet of the chamber. Note that the measurement is stable across the 90 minute recording. (b) We correct for the time lag between sensors and plot the mapping between PID measurements and OS readings.}{\includegraphics[width=10cm]{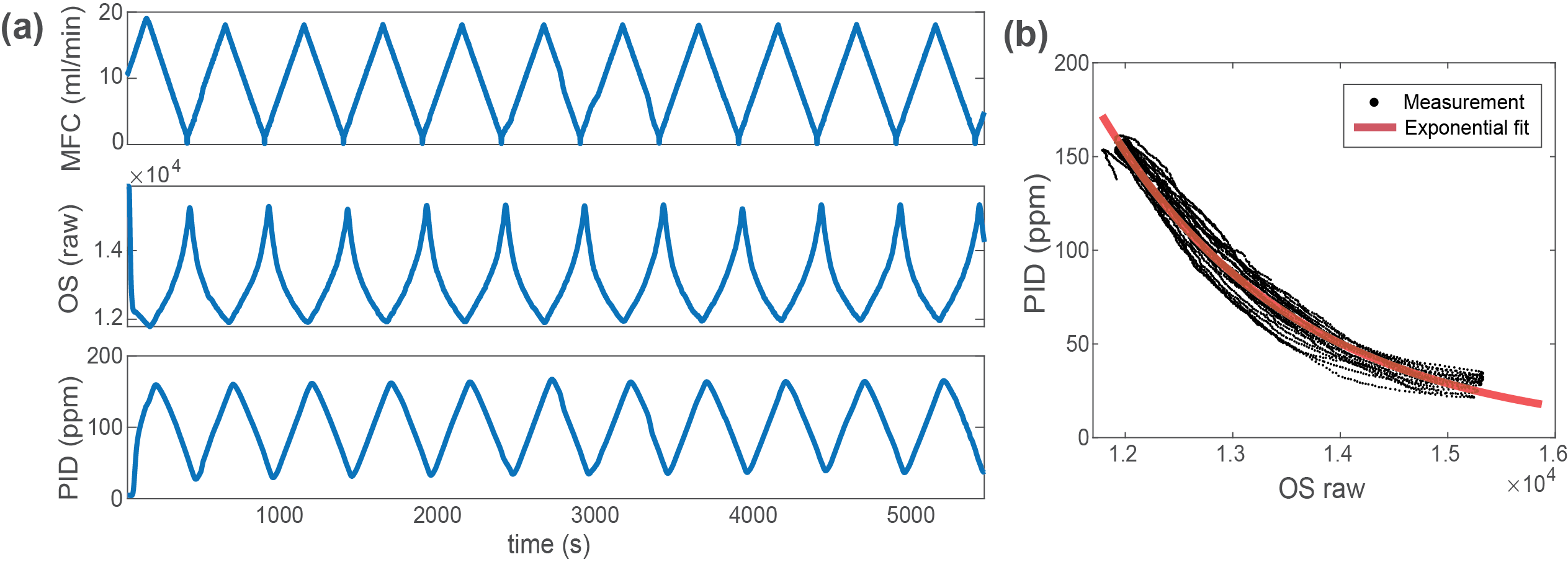}}
\label{figsupp:figSI1-2}
\end{fullwidth}
\end{figure}

We constructed a flow chamber to control odor air flow across an arena (\FIG{fig1}a). Odor and humidified air are sourced from two bubblers, one containing pure water and the other an aqueous solution of odorant and water. Flow rates are controlled by separate mass flow controllers (MFCs) upstream of the bubblers. Downstream of the bubblers, the odor and air streams are divided into parallel sections of equal lengthed tubing. Each tube is connected to one input port of the flow chamber. The pattern of connections and the flow rates set by the two MFCs determines the shape of the produced odor profile. For instance, if odor is provided at a single central inlet, the resulting profile is a `cone' (\FIG{fig1}d-g) whose peak concentration and divergence are controlled with the MFCs (e.g. speeding up the odor flow while slowing down the air flow broadens the cone). Temporal gradients can be achieved by varying the odor flow in time, subject to constraints imposed by the odor's absorption into the agar gel.

The outflux from the flow chamber is connected to a flow meter and photo-ionization detector to monitor the overall flow rate and odor concentration respectively. The geometry of this flow chamber is shown in (\FIG{fig1}b), where parallel tubings are connected from the side and the chamber is vacuum sealed with a piece of acrylic on top during experiments. The chamber is designed for use with $\sim$100 mm square agar plates. The extra width (2.5 cm on either side) diminishes the influence of the chamber boundary on the odor profile over the arena. Interchangeable inserts allow for different agar substrates (e.g. circular plates) or for full calibration by odor sensor arrays (\FIG{fig1}c), discussed in the next section. Metal components are designed for low-cost fabrication by automated mechanisms (either laser cut-able or 3-axis CNC machinable). The fabrication plans for the flow chamber, design for the agar plate inserts, and required components to construct the flow path are publicly available in the \nameref{ssec:num1} section.

\subsection{Measuring the spatiotemporal odor distributions} A central difficulty in measuring animals' responses to olfactory cues is quantifying airborne odor concentrations that vary in space and time. This difficulty is exacerbated in turbulent environments where odor plumes carry abrupt spatial and temporal jumps in concentration far from the source with fundamentally unpredictable dynamics. But even in laminar flows, boundary conditions, slight changes in temperature, and the presence of absorbing substrates like agar make this challenging. 

There is therefore a need, even for quasi-stationary gradients, to characterize the odor profiles in situ and to monitor these profiles during experiments. Various optical techniques, like laser induced fluorescence or optical absorption \citep{Louis2008-ju,tadres2022depolarization,demir2020walking}, %
exist to monitor concentration across planar arenas, but in general, these are incompatible with behavior experiments, expensive to construct, require specially designed arenas, or some combination of these disadvantages. Electronic chemical sensors can reveal the time-varying concentration at a particular point in space. A tiled array of these sensors acts as a `camera' forming a 2D spatiotemporal reading of the concentration. The gold-standard for measurement of odor concentration is the photo-ionization detector (PID), but even the smallest versions of these sensors are both too large ($\sim$ 2 cm in all dimensions) and too expensive ($\sim$ \$500 each) to make an array. Metal-oxide odor sensors, designed to be used in commercial air quality sensors, are available in inexpensive and compact integrated circuit packages. However, in general, commercial metal-oxide sensors are not designed for precision work - they tend to drift due to variations in heater temperature, humidity, adsorption of chemicals and ageing effects. Most such sensors are designed to detect the presence of gas above a particular concentration but not to precisely measure the absolute concentration. We became aware of a newer metal-oxide sensor, the Sensirion SGP30 that was designed for long-term stability and concentration measurement; we wondered if such a sensor could be calibrated for use in an odor sensor array.

To calibrate the sensor, we created a controllable concentration source by bubbling air through butanone. The odor reservoir contains butanone dissolved in water and is kept below the saturation concentration (11 mM or 110 mM odor sources).
We then mixed this odorized air flow into a carrier stream of pure air. We kept the carrier air flow rate constant ($\sim 400$ mL/min) and varied the flow rate through the odor source ($0-50$ mL/min); the odor flow rate was slow enough that the vapor remained saturated, so the concentration of butanone in the mixed stream was proportional to the flow rate through the butanone bubbler, as directly measured with a PID (\FIGSUPP[fig1]{figSI1-2}). We typically calibrated concentration with continuously ramped flow rate in triangle wave with 500 s period for 2-3 cycles.
We found a one-to-one correspondence between the odor sensor reading and the PID reading that persisted over time and showed no hysteresis. We reasoned that after applying this calibration procedure to an array of sensors, we could use the array to measure spatiotemporal odor concentration distributions with accuracy derived from the PID. Continuous calibration for 90 minutes showed that the odor sensors reliably reported concentration across durations (\FIGSUPP[fig1]{figSI1-2}) much longer than the typical behavioral experiment.

We constructed the sensor array from `odor sensor bars' (OSBs), printed circuit boards each containing 16 sensors in two staggered rows of 8. Each OSB also contained 8 temperature and humidity sensors to allow compensation of the odor sensor readings. The OSBs are mounted orthogonal to the direction of air flow; 7 OSBs fit inside our flow chamber (112 sensors total) allowing a full measurement of the odor profile. Taken together these 112 sensors formed an odor sensor array (OSA), capable of measuring odor concentrations with %
$\sim$ 1 cm spatial and 1 second temporal resolution. Prior to all experiments we calibrated the OSA in situ by varying the butanone concentration across the entire anticipated range of measurement while simultaneously recording the odor sensor and PID readings. 

To verify the ability of the OSA to measure concentration gradients we created an artificially simple steady-state odor landscape by flowing odorized air ($\sim 30$ mL/min) through a central tube and clean air through the others ($\sim 400$ mL/min distributed into 14 surrounding tubings) in an environment without agar and without animals. This results in an air flow velocity $\sim 5$ mm/s in the flow chamber.  As the flow rates and concentrations are all known and the flow is non-turbulent, and there is no agar or animals present, the concentration across the chamber should match a convection-diffusion model. After establishing the gradient, we recorded from the discrete odor sensors on the array (\FIG{fig1}d) and estimated the values in between sensors using spline interpolation with length scale equal to the inter-sensor distances (\FIG{fig1}e). We compared this stationary profile with a two-parameter convection-diffusion flow model (\FIG{fig1}f,g) fit to the data and described in the Methods section. 
The measured concentrations in this artificially simplistic odor gradient show good agreement with the fit convection-diffusion model, especially in the central region where experiments are to be conducted, leading us to conclude the OSA can accurately report the odor concentration. We proceed to consider more complex odor environments. %

\begin{figure}
\begin{fullwidth}
\includegraphics[width=1.0\linewidth]{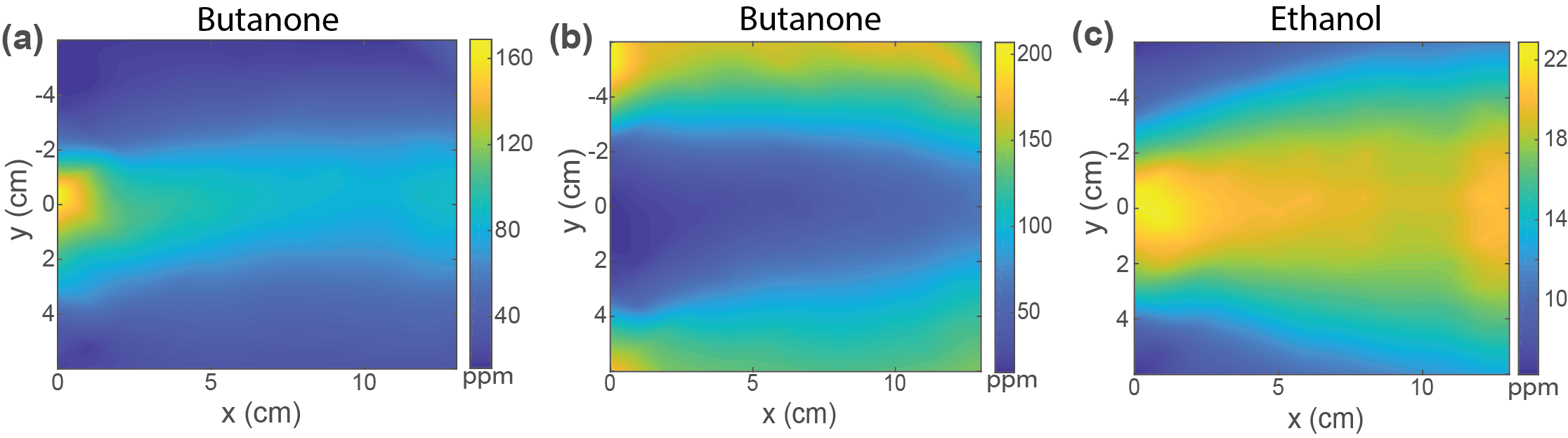}
\caption{
\textbf{Flexible control of a steady-state odor landscape.} \textbf{(a)} Configuration for a narrow cone with butanone. Colorbar shows interpolated odor concentration in ppm as measured by the odor sensor array. \textbf{(b)} An inverse cone landscape that has higher concentration of butanone on both sides and lower in the middle. \textbf{(c)}Another stationary odor landscape of a different shape, this time with ethanol.
\label{fig:fig2}
}
\end{fullwidth}
\end{figure}

We demonstrate several examples of flexible control over the odor profile.  In the configuration shown in \FIG{fig1} the center-most input provides odorized air and all surrounding inputs provide moisturized clean air to form a cone-shape stationary odor pattern. A narrower cone can be created by increasing the air flow (600 mL/min) of the surrounding inputs relative to the middle odorized flow (\FIG{fig2}a). %
The cone can be inverted by placing odorized air in the 
two most distal inputs, and clean air in all middle inputs (\FIG{fig2}b). This inverse cone has lower concentration in the middle and higher on the sides. In later animal experiments, we restrict odorized air to one side to form a biased-cone odor landscapes, resulting in a cone with an offset from the middle line of the arena. Many more configurations are possible, demonstrating that the odor flow chamber enables the flexible control of airborne odor landscapes that are much more complex than a single odor point source. To show that the flow control and measurement methods are not restricted to any single odor molecule, we created and measured a cone profile using ethanol (\FIG{fig2}c).

\subsection{Odor-agar interactions dominate classical droplet assays}

\begin{figure}
\begin{fullwidth}
\includegraphics[width=1.0\linewidth]{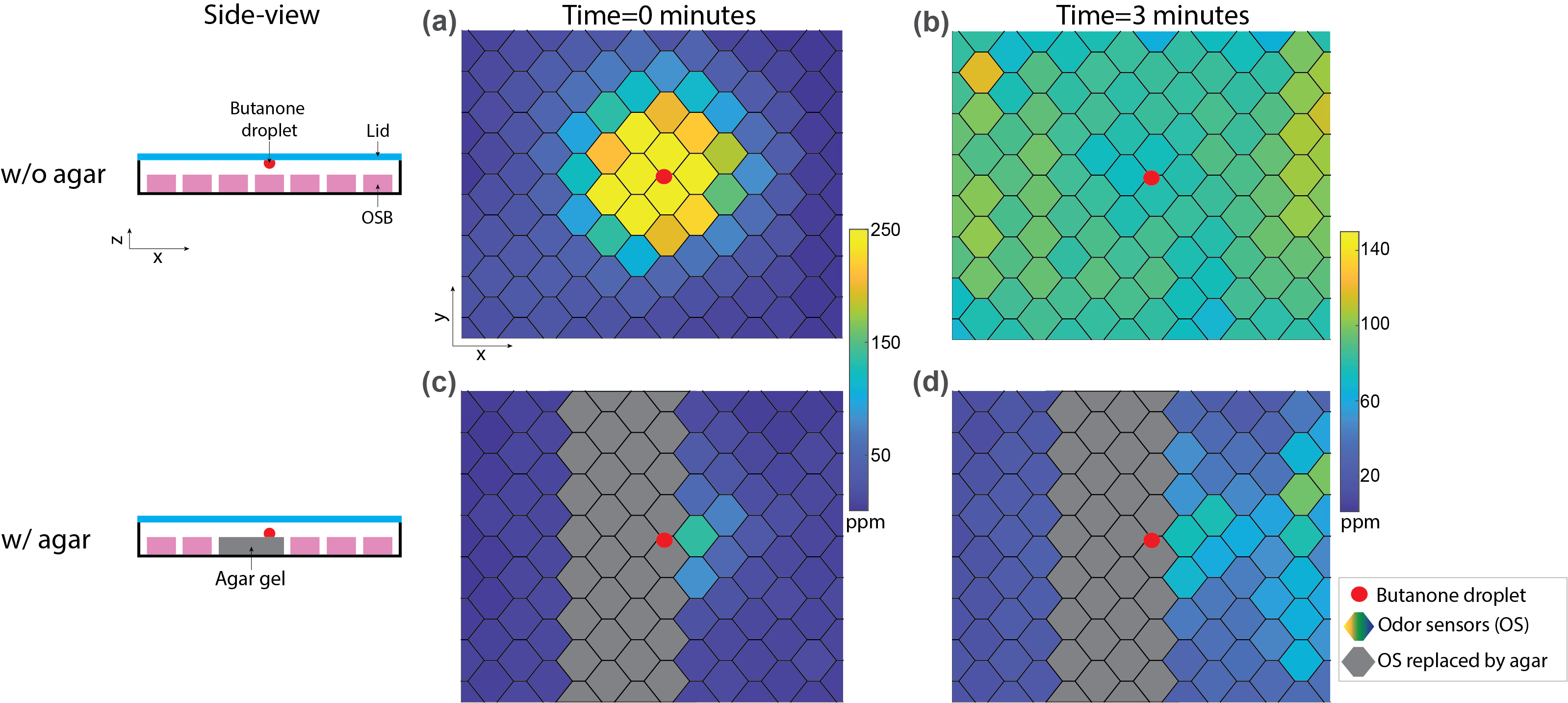}
\caption{\textbf{The presence of agar in the droplet assay alters the time-evolution of butanone odor landscape.}
\textbf{(a)} Concentration measured by odor sensor array is reported immediately after butanone droplet is introduced into the arena without agar. Red dot indicates the position of butanone droplet on the lid over the sensor array ($2\mu L$ of $10\%$ v/v butanone in water). \textbf{(b)} Same but three minutes later.
\textbf{(c)} Same measurement as in (a) but now droplet is added onto agar (gray). Two odor sensor bars have been removed to make space for the agar. \textbf{(d)} Same as (c) but three minutes later. Side view of the configurations with OSB sensors, butanone droplets, and agar gel are shown on the right.}
\label{fig:fig3}
\figsupp[Concentration measurements with an odor droplet and experimental perturbation.]{Concentration measurements with an odor droplet and experimental perturbation. (a) The initial concentration readout from a droplet of butanone on the lid near the middle of the arena. (b) Same condition as (a), but 20 minutes after the recording (left) and after shortly opening and closing the lid back (right) to mimic perturbation during worm experiments. (c) When there is agar in the chamber, the odor concentration is better maintained in the chamber. Location of the odor droplet is shown with a red dot. (d) After 20 minutes of recording (left) and after shortly opening and closing the lid back (right) to mimic experimental perturbation.}{\includegraphics[width=10cm]{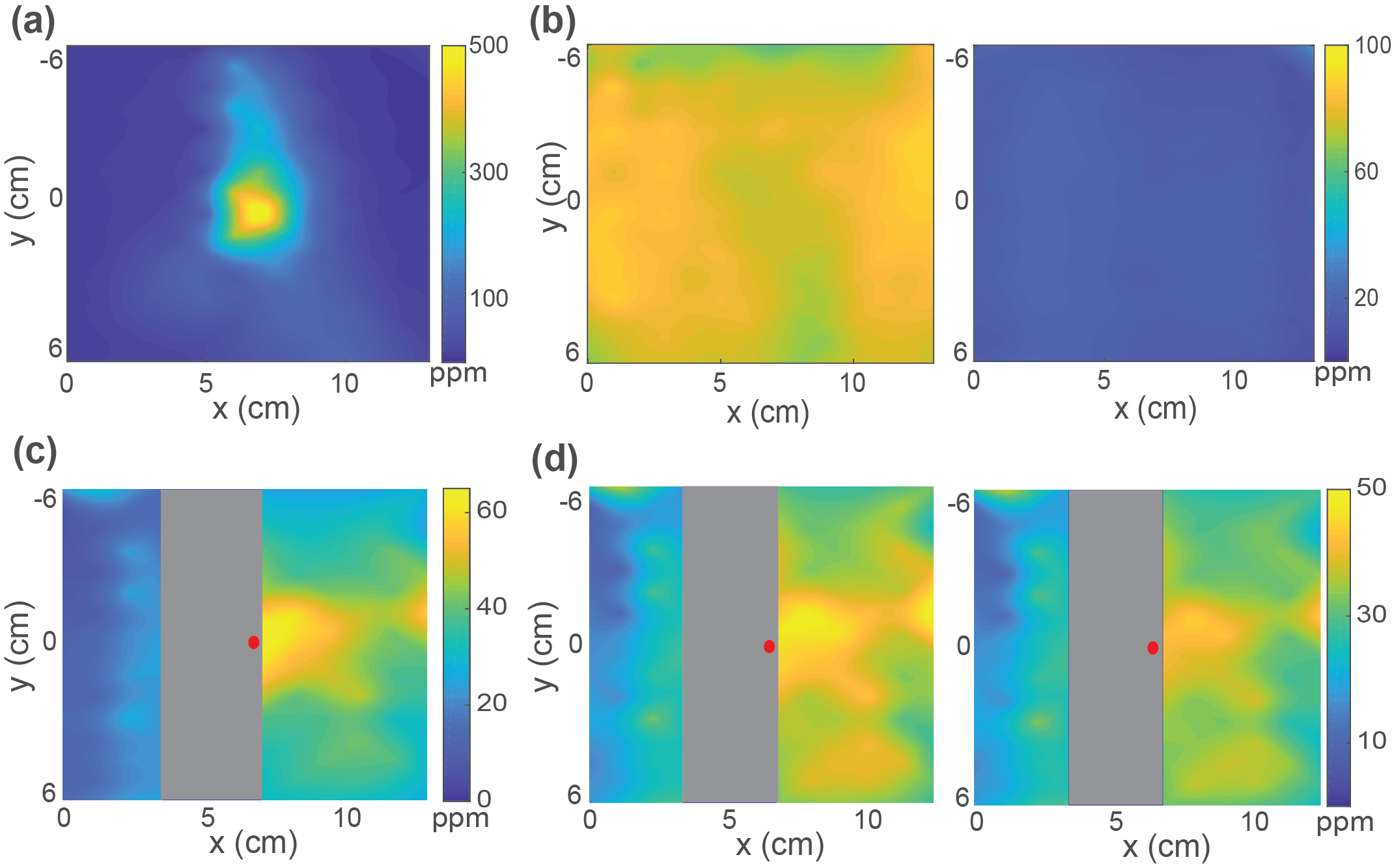}}\label{figsupp:figSI3}
\videosupp{Time-evolution of odor landscape from a butanone droplet with (right) and without (left) agar. The experimental conditions are the same as \FIG{fig3}. Blank sensor positions indicate sensors replaced by agar. The video updates every 2 seconds and measures butanone concentration from a droplet in the first 3 minutes. 
Video available online at  \href{https://figshare.com/articles/dataset/Continuous_odor_profile_monitoring_to_study_olfactory_navigation_in_small_animals/21737303}{10.6084/m9.figshare.21737303}
}\label{videosupp:sv1}
\end{fullwidth}
\end{figure}

Classic chemotaxis experiments in small animals commonly construct odor environments with odor droplets in a petri dish, usually with a substrate like agar. Our odor delivery instrument is designed to be compatible with a similar environment. To first better understand classical chemotaxis experiments, we sought to characterize the spatiotemporal odor profile from an odor droplet point source using our odor sensor array. We first considered the case without agar. In that case the odor concentration should be governed entirely by gas-phase diffusion. %
We placed a $2 \mu$L droplet of $10\%$ butanone in water %
on the lid of our instrument centered in the arena above the full OSA and without any airflow (\FIG{fig3}a,b). Butanone was observed to diffuse across the arena in the first three minutes (supplementary video (\VIDEOSUPP[fig3]{sv1})
and the equilibrium concentration is close to uniform across the odor sensors. We note that the final concentration of roughly 100 ppm, and the equilibration timescale both match what we would expect from first principals for $\sim 10^{-6}$ mol of butanone in a $\sim 225 \text{mL}$ arena that have a diffusion rate of $\sim 0.08 \text{cm}^2/$s in air.  
A uniform odor landscape is not helpful for studying odor guided navigation, but most behavioral experiments are not conducted in a bare flow chamber but contain a biologically compatible substrate, such 
as an gar gel, as is typically used in droplet assays. We therefore sought to investigate the role that agar plays in sculpting the odor landscape.

We introduced agar into the droplet assay by removing two sensor bars and replacing them with agar. We placed a butanone droplet directly on the agar, as done classically, and measured the odor landscape over time (\FIG{fig3}c,d). The odor concentration measured with agar is dramatically different from that measured without agar. Instead of quickly equilibrating to a uniform concentration, in the presence of agar there was instead a local maximum of butanone surrounding the droplet that persists even after 3 minutes. This difference in airborne odor concentration with and without agar persists after experimental perturbation such as removing and replacing the lid over the chamber (\FIGSUPP[fig3]{figSI3}). More broadly, the odor landscape we observed in the presence of agar would have been hard to predict ahead of time. An important consequence of this finding is that, to create a specific odor landscape (as in \FIG{fig1} or \FIG{fig2}) with agar, one will need to account for the effect of agar. We therefore sought to study odor-agar interactions more systematically and in the context of air flow. %

\subsection{Measuring and compensating odor-agar interactions with flow}

We first sought to measure whether the presence of agar changed the odor profiles generated due to flow in a bare chamber (\FIG{fig1}, \FIG{fig2}). As in \FIG{fig3}c,d, we replaced two odor sensor bars with a rectangular strip of agar gel or a metal plate as a control, and then measured airborne odor concentration upstream and downstream of the agar under odorized airflow that would normally produce a cone profile (\FIG{fig4}a-b).
While the agar had little effect on the odor landscape upstream of the agar, it drastically altered the downstream odor landscape (\FIG{fig4}b), suggesting that the agar absorbs
the airborne butanone molecules. This finding is consistent with the odor droplet experiments (\FIG{fig3}) and to be expected since butanone is highly soluble in water (275 g/L). Pulse-chase style experiments confirm that agar does indeed absorb and reemit butanone (\FIGSUPP[fig4]{figSI4-2}b). We also observed disruptions to the odor landscape when we used a full-sized 96 mm square agar plate intended for use with animals, \FIG{fig4}c. To accommodate the full sized agar plate we measured only the one dimensional odor profiles upstream and downstream of the agar \FIG{fig4}d. Taken together, these experiments suggest that agar-butanone interaction presents a challenge for setting up and maintaining stable odor landscapes. 

We next sought a method to generate desired odor landscapes even in the presence of agar. We generate the odor profile by constant flow, which continuously replenishes the airborne odor. In principle, the disruption caused by agar should be overcome by constant flow of a sufficiently long duration, after which the agar and airborne odor would be in quasi-equilibrium at all spatial locations, with the concentration of odor dissolved in the gel proportional to the airborne concentration above it. 

We measured odor concentration downstream of the agar and found that the airborne concentration failed to approach equilibrium on the timescales of single experiments \FIG{fig4}e,f. This suggests that it is not practical to simply wait for the agar and odor to reach equilibrium. Instead we developed a pre-equilibration protocol to more efficiently bring the agar and airborne odor into equilibrium before our experiments.
    
To more rapidly establish a desired airborne odor landscape, we briefly first exposed the agar to an airflow pattern corresponding to higher-than-desired odor concentration, created by replacing the odor reservoir with one containing a higher concentration of butanone.
We monitored the odor profile downstream of the agar until it reached the desired concentration and then switched to the original bubbler to maintain that concentration. Using this pre-equilibration protocol, we reached quasi-equilibrium quickly, typically after the order of ten minutes, \FIGSUPP[fig4]{figSI4-2},c. Note the spatial parameters of the two airflow patterns were the same, only the concentration of the odor source changes. Pre-equilibration allows the generation of airborne odor gradients in the presence of agar that match those in the absence of agar (\FIG{fig4}b,d,f right vs left column). We modeled the pre-equilibration protocol using a reaction-convection-diffusion model considering first order interactions between odor and agar. Under reasonable assumptions about the absorption rate, reemission rate, and capacity of the agar, simulations of this simple model provided qualitative agreement to our observations (\FIG{PE_model}a-c).

\subsection*{Monitoring the boundary determines the odor landscape at quasi-equilibrium}

\begin{figure}
\begin{fullwidth}
\includegraphics[width=0.95\linewidth]{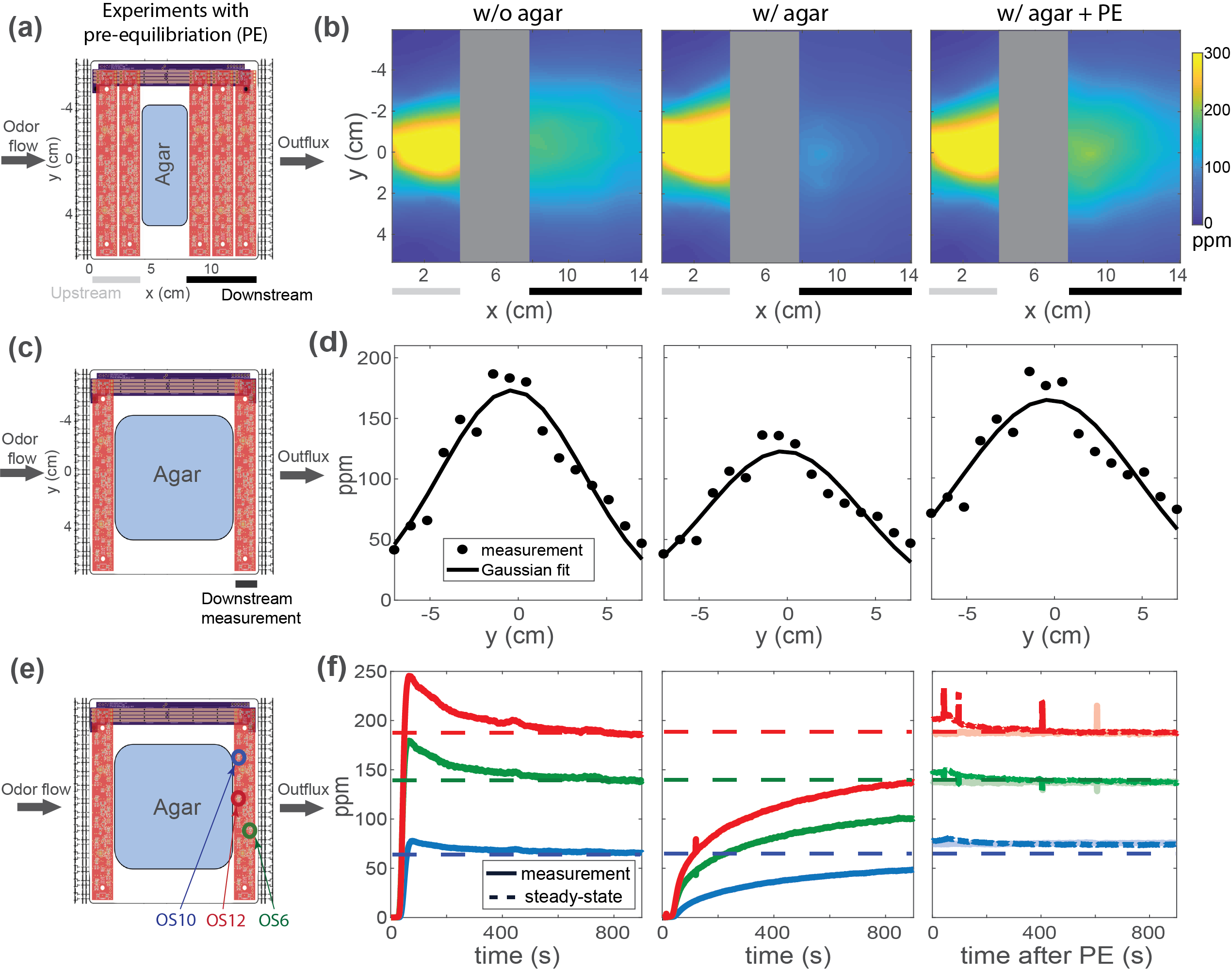}
\caption{\textbf{Under flow, agar interacts with odor to disrupt the downstream spatial odor profile, but a  pre-equilibration protocol can coax the system into quasi-equilibrium and restore the odor profile.} %
\textbf{(a)} Two odor sensor bars are replaced with agar to observe the effect of introducing agar on the downstream spatial odor profile. %
\textbf{(b)} Measured odor profile is shown upstream and downstream of the removed odor sensor bars in the absence (left) and presence of agar (middle).
Transiently delivering a specific higher odor concentration ahead of time via a pre-equilibration (PE) protocol restores the downstream odor profile even in the presence of agar. \textbf{(c)} Additional odor sensor bars are removed and replaced with a larger agar, as is typical for animal experiments. \textbf{(d)} Measurements from the downstream sensor bar under the same three conditions in (b). The dots are sensor measurements and the smooth curve is a Gaussian fit. \textbf{(e)} The same experimental setup in (a), here focusing on time traces of only three downstream odor-sensors (colored circles for selected OS). \textbf{(f)} Concentration time series of three sensors color-coded in (e). Traces for three conditions are shown: time aligned to initial flow without agar (left), time aligned to initial flow with agar (middle), and traces after PE (right, with transparent line showing measurements another 20 min after the protocol). The dash-lines indicate the target steady-state concentration for each sensor.
\label{fig:fig4}
}
\figsupp[Time series of concentration change that capture effects of agar gel and the PE protocol]{Time series of concentration change that capture effects of agar gel and the PE protocol. (a) Concentration readout from the downstream PID (top) in response to the impulse of air flow rate though the odor bottle controlled by MFC (bottom), with no agar in the flow chamber. The background clean air flow is constant $\sim$ 400 mL/min throughout the recording. (b) Same as (a) but with agar plate in the flow chamber. Note that the response time scales to the same impulse are significantly different. (c) The time trace recorded from the PE protocol with agar plates. Odor reservoir with high butanone concentration (110 mM) is applied in the beginning, swapped back to the target concentration (11 mM) at $\sim$ 200 seconds, the odor concentration readout relaxes and stabilizes after $\sim$ 900 seconds, which enters steady-state for a duration longer than animal experiments.
}{\includegraphics[width=10cm]{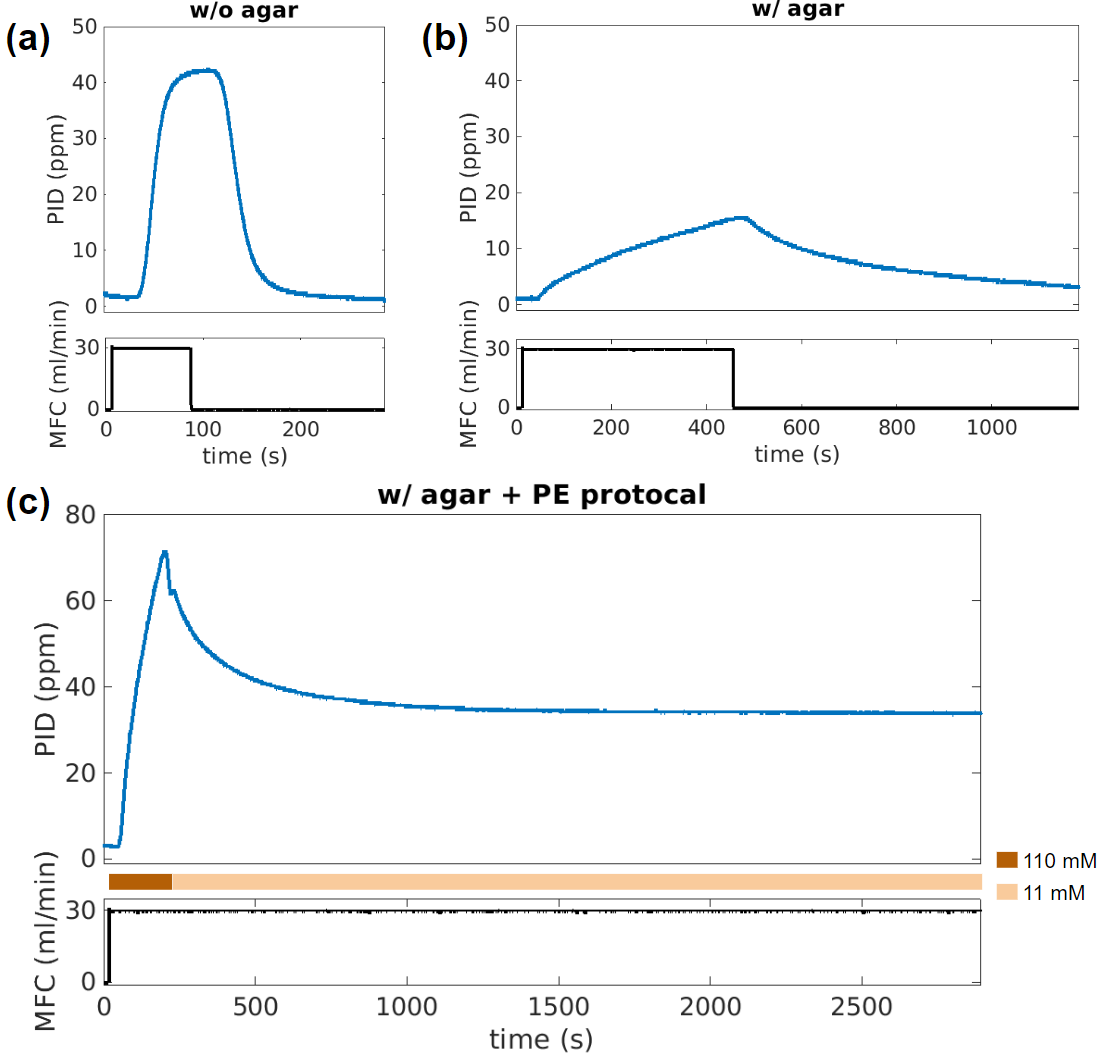}}
\label{figsupp:figSI4-2}
\end{fullwidth}
\end{figure}

\begin{figure}
\begin{fullwidth}
\includegraphics[width=0.9\linewidth]{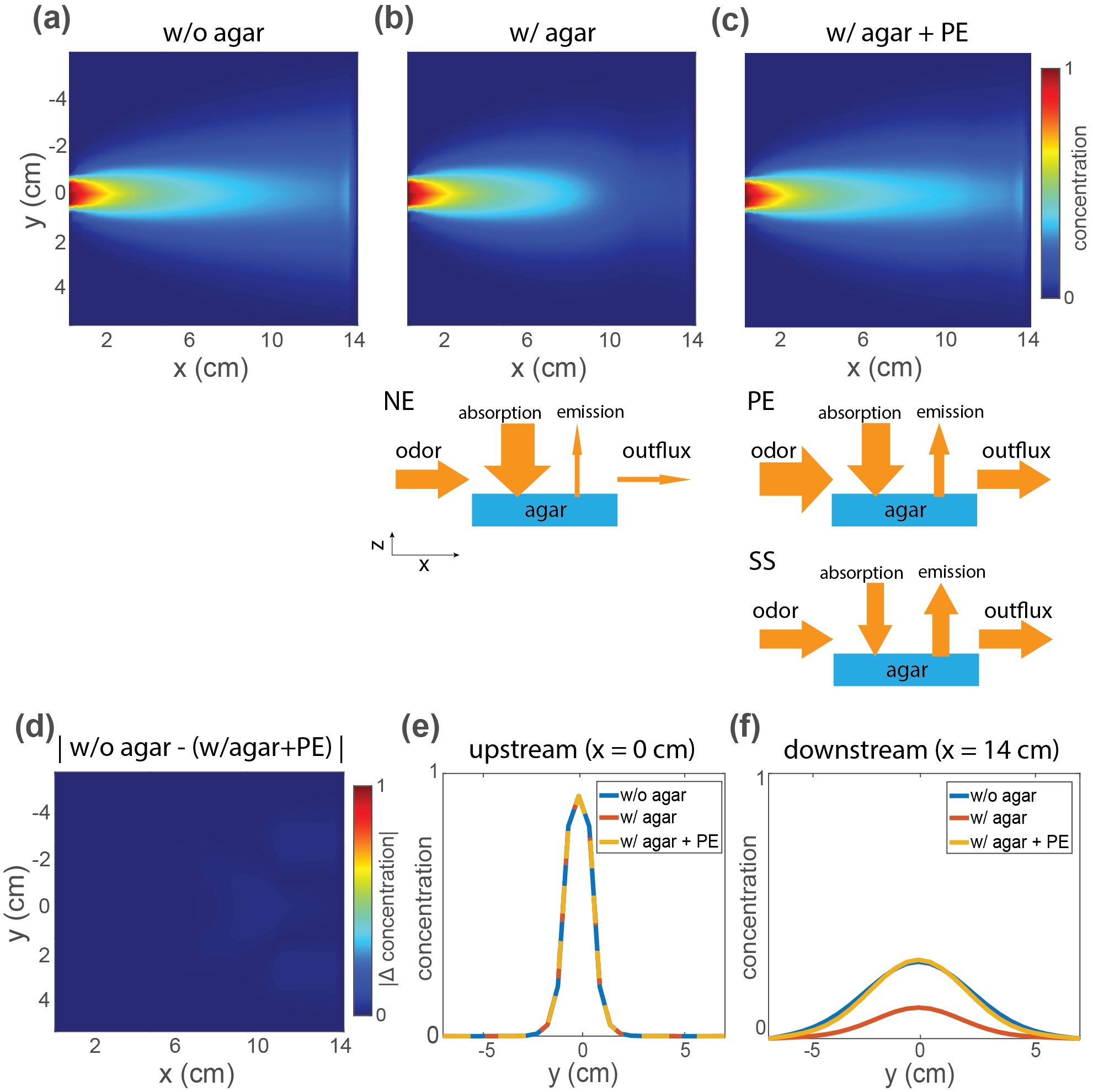}
\caption{\textbf{Simulations from a reaction-convection-diffusion model of  odor-agar interaction show that at quasi-equilibrium the airborne odor concentration is the same with or without agar.} \textbf{(a)} Simulation results of steady-state odor concentration in air without agar and with  flow configured as in \FIG{fig1}. \textbf{(b)} The same simulation condition in (a) but now shortly after agar is introduced and before quasi-equilibrium is reached. A schematic of odor-agar interaction model is shown below. When agar is introduced, it absorbs the odor in air and decreases concentration measured downstream, producing a non-equilibrium (NE) concentration profile. \textbf{(c)} Odor concentration profile in air, with agar present, but after the pre-equilibration (PE) protocol brings this system to quasi-equilibrum. The PE protocol is shown in the schematic below, followed with steady-state (SS) with the stable odor concentration profile shown above. \textbf{(d)} The absolute difference of concentration profile without agar (a) and with agar after PE (c) is shown. \textbf{(e)} Upstream and \textbf{(f)} downstream odor concentrations along the agar boundary are shown for all three conditions.
    }
    \label{fig:PE_model}
    \end{fullwidth}
\end{figure}

It is critical to monitor the odor landscape during animal experiments because the landscape is sensitive to environmental and experimental conditions which may fluctuate within and between experiments. But it is inconvenient to measure airborne concentration directly over the agar (e.g. because sensors impede optical access and also require heat management). 
Fortunately, measuring the odor profile upstream and downstream of the agar places strong constraints on the airborne odor concentration over the agar such that in practice the spatial concentration can be confidently inferred.

If the airborne odor concentration upstream and downstream of the agar matches the profile in the absence of agar, one can infer that the airborne odor concentration landscape above the agar is also the same. The argument is straightforward: in the absence of sources or sinks, the fact that two concentration distributions obey the same differential equations and share the same conditions on all boundaries means that the the distributions are identical throughout the interior. Given identical measurements of the with-agar and without agar profiles at the inlet and outlets and reflecting boundary conditions on both walls, the only way for the with-agar distribution to differ from that without-agar is for sources and sinks of odor in the agar to be precisely arranged so that the all excess odor emitted from one point is exactly reabsorbed somewhere else before reaching the boundary. Not only is such an arrangement unlikely, it is inherently temporally unstable.  A mathematical version of this argument is presented in the \hyperref[ssec:appendix]{Appendix}.

This quasi-equilibrium argument is supported by empirical concentration measurements shown in \FIG{fig4} and numerical results with the reaction-convection-diffusion model demonstrated in \FIG{PE_model}. The simulation results show that there is negligible difference between conditions with and without agar at quasi-equilibrium (\FIG{PE_model}d) when the odor concentration along the boundary  is the same (\FIG{PE_model}e,f). Together, our numerical estimations, along with empirically observations allow us to safely infer that when measurements along the boundary indicate that they system is in quasi-equilibrium, the odor concentration experienced by animals on agar are the same as the concentrations measured in the absence of agar.

\subsection{Butanone chemotaxis in \emph{C. elegans}} %
We sought to directly quantify \textit{C. elegans}' navigation strategies for airborne butanone using our odor delivery system.
\textit{C. elegans} are known to climb gradients towards butanone \citep{bargmann1993odorant, Cho2016-is, Levy2020-oh}. Microfluidic environments suggest that they use a biased random walk strategy to navigate in a liquid butanone environment \citep{Levy2020-oh, Albrecht2011-fj}. Worms are also known to use weathervaning to navigate airborne odor gradients \citep{Iino2009-al, kunitomo2013concentration} although to our knowledge this has not been specifically investigated for butanone.

\begin{figure}
\begin{fullwidth}
\includegraphics[width=1.0\linewidth]{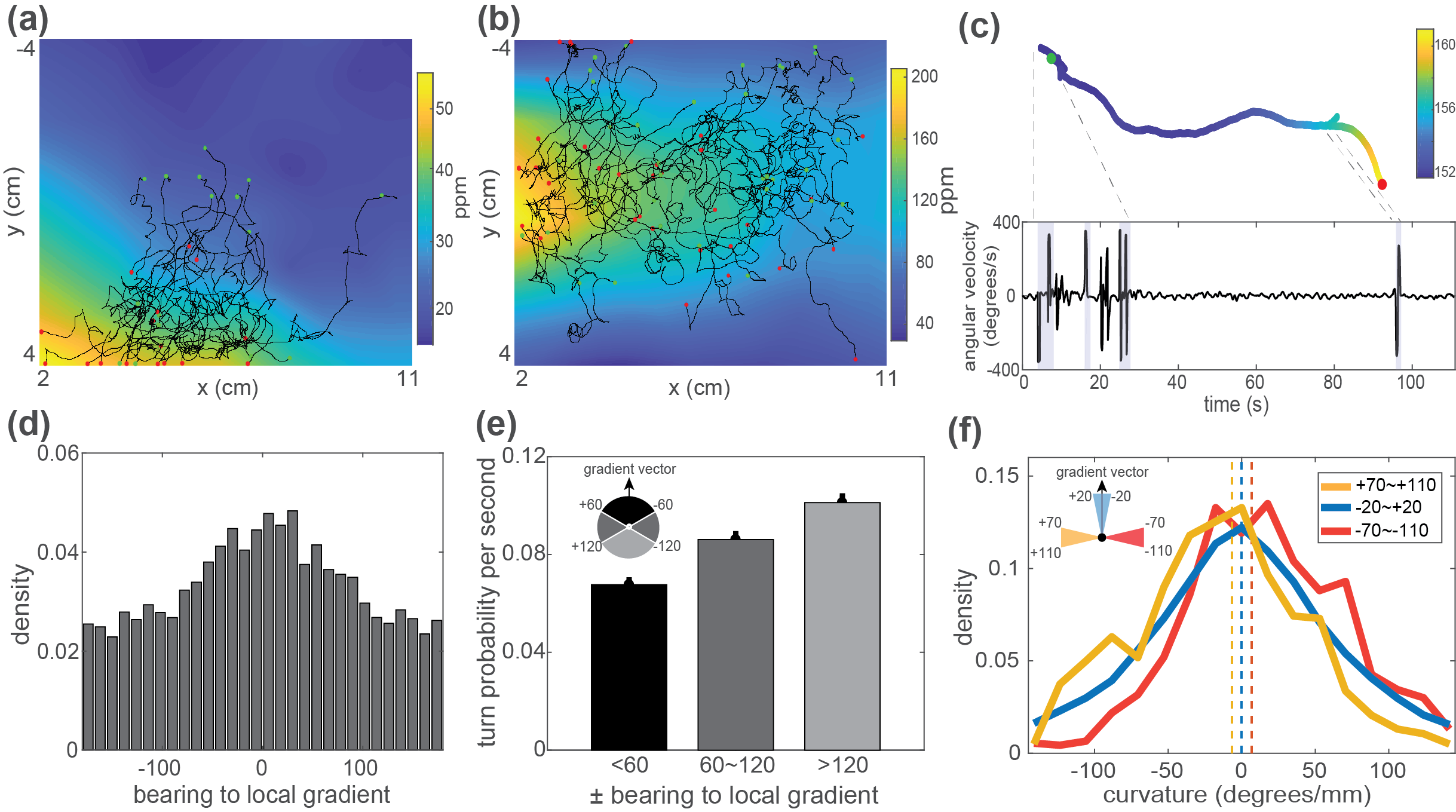}
\caption{\textbf{\textit{C. elegans} use both biased random walk and weathervaning to navigate in a butanone odor landscape.} Animals on agar were exposed to butanone in the flow chamber.  \textbf{(a,b)} Measured animal trajectories are shown overlaid on airborne butanone concentration for different odor landscapes. Green dots are each animal’s initial positions and red dots are the endpoints. %
\textbf{(c)} An animal's trajectory is shown colored by the butanone concentration it experiences at each position (top). Its turning behavior is quantified and plotted over time. Turning bouts are highlighted in gray. \textbf{(d)} Distribution of the animal's bearing with respect to the local airborne odor gradient is shown. Peak around zero is consistent with chemotaxis. \textbf{(e)} 
Probability of observing a sharp turn per time is shown as a function of the absolute value of the bearing relative to the local gradient. Modulation of turning is a signature of biased random walk. Error bars show error for counting statistics. Data analyzed from over 9,000 tracks produced from $\sim$300 worms, resulting in 108 hours of observations. \textbf{(f)} Probability density of the curvature of the animal's trajectory is shown conditioned on bearing with respect to the local gradient. Weathervaning strategy is evident by a skew in the distribution of trajectory curvature when the animal travels perpendicular to the gradient (yellow and red). Three distributions are significantly different from each other according to two-sample Kolmogorov-Smirnov test ($p < 0.001$). Means are shown as vertical dashed lines.
\label{fig:fig5}}
\figsupp[Control measurements with air flow and without odor gradient.]{Control measurements with air flow but no odor. (a) Behavioral trajectories overlaid on the odor landscape that would have been expected had odor been present (mock odor landscape). No odor is presented, only  moisturized airflow. (b) Distribution of bearing to the mock gradients under clean air flow. (c) Turn probability at different  bearing conditions to the mock gradient. (d) Curvature conditioned on different  bearing measurements.
}{\includegraphics[width=10cm]{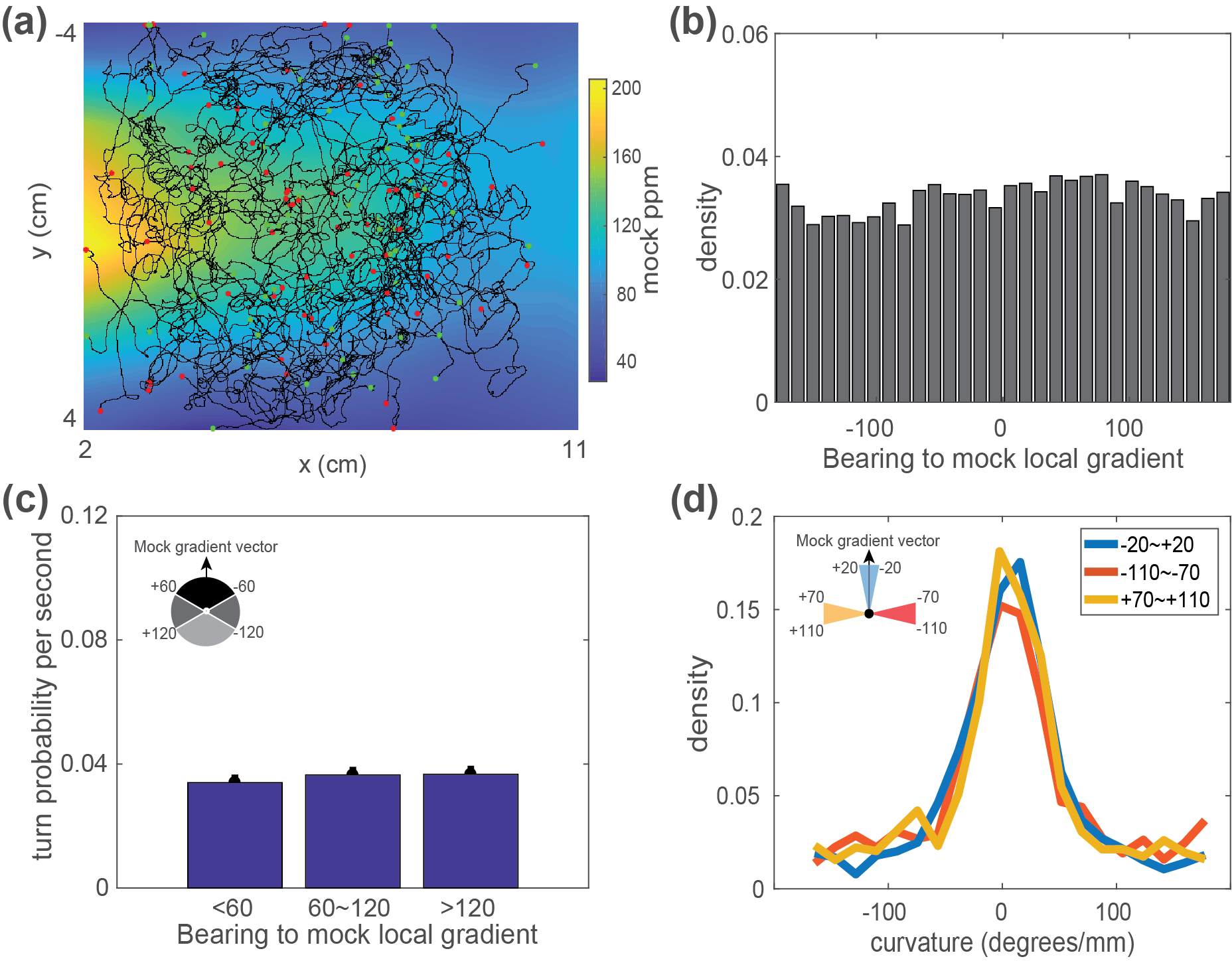}}
\label{figsupp:no_odor_control}
\end{fullwidth}
\end{figure}

Worms were imaged crawling on agar in the flow chamber under an airborne butanone odor landscape illuminated by infrared light. Here 6 recording assays were presented, with approximately $50-100$ animals per assay, and two different odor landscapes were used.
\textit{C. elegans} navigated up the odor gradient towards higher concentrations of butanone, as expected (\FIG{fig5}a,b). Importantly, the odor concentration experienced by the animal at every point in time was inferred from concurrent measurements of the odor profile along the boundary of the agar, \FIG{fig5}c. On average, animals were more likely to travel in a direction up the local gradient than away from the local gradient, as expected for chemo-attraction \FIG{fig5}d. We use the term ``bearing to local gradient'' to describe the animal's direction of travel with respect to the local odor gradient that it experiences.

We find quantitative evidence that the worm exhibits both biased random walk and weathervaning strategies. 
To investigate biased random walks, we measured the animal's probability of turning (pirouette) depending on its bearing with respect to the local airborne butanone gradient \FIG{fig5}e.
We find that the animal is least likely to turn when it navigates up the local gradient and most likely to turn when it navigates down the gradient, a key signature of the biased random walk strategy \citep{berg2018random, mattingly2021escherichia}.  

To test for weathervaning, we measured how the curvature of the animal's trajectory depended on its bearing with respect to the local airborne butanone gradient, \FIG{fig5}f. 
When the animal navigated up the butanone gradient (\FIG{fig5}f, blue) the distribution of the curvature of its trajectory was roughly symmetric and centered around 0 (straight line trajectory). By contrast, when the animal navigated perpendicular to the gradient (\FIG{fig5}f, yellow and red) the distribution of the curvature of its trajectories was skewed. The skew was such that it enriched for cases where the animal curved its trajectories towards the local gradient, a key signature of weathervaning.
Both the biased random walk and weathervaning behavior was absent in control experiments with flow but not odor, and we observed no evidence of anemotaxis at the $\sim 5$mm/s air velocities encountered by the animals   (\FIGSUPP[fig5]{no_odor_control}).
We conclude that \textit{C. elegans} utilize both biased random walk and weathervaning strategies to navigate butanone airborne odor landscapes. We note that the quantitative analysis needed to make this conclusion relied on knowledge of the local airborne odor gradient experienced by the animal, which was provided by our odor profile measurements.

\begin{figure}
\begin{fullwidth}
\includegraphics[width=0.7\linewidth]{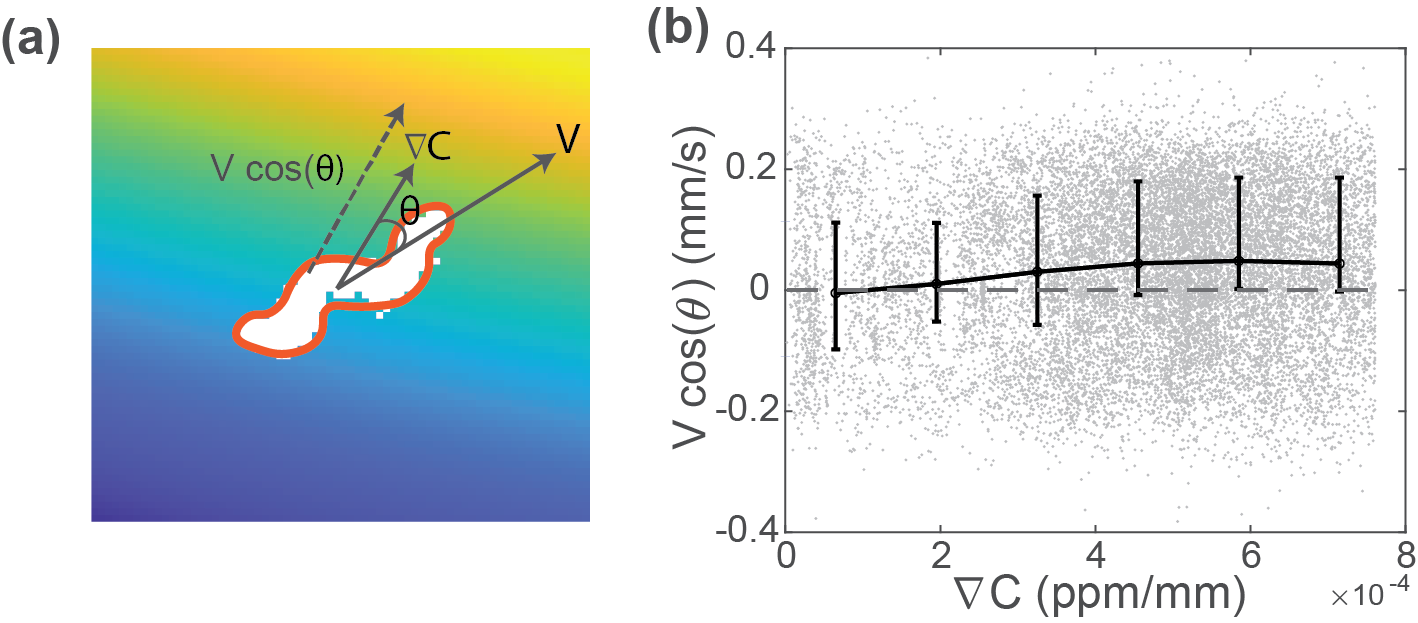}
\caption{
\textbf{Tuning curve relating animal drift velocity to experienced odor concentration gradient.} \textbf{(a)} Schematic of a worm tracked in the odor landscape. The crawling velocity vector $V$, local concentration gradient $\nabla C$, bearing angle $\theta$, and drift velocity $V\cos(\theta)$ are shown. \textbf{(b)} Tuning curve shows the drift velocity $V\cos(\theta)$
as a function of the odor concentration gradient. Gray dash line indicates an unbiased performance with zero drift velocity, gray dots are the discrete measurements, and the black line shows the average value within bins. Error bar shows lower and upper quartiles of the measurements. %
}
\label{fig:fig6}
\end{fullwidth}
\end{figure}

To quantify the overall navigational response with respect to local gradients, we further compute the animal's drift velocity as a function of local gradients (\FIG{fig6}). This captures the animal’s overall gradient climbing performance as a result of all the navigational strategies it uses, including the biased random walk and weathervaning. This calculation is only possible with a knowledge of the odor concentration experienced by the animal.

\subsection{Butanone chemotaxis in \emph{Drosophila} larvae}
\begin{figure}
\begin{fullwidth}
\includegraphics[width=0.9\linewidth]{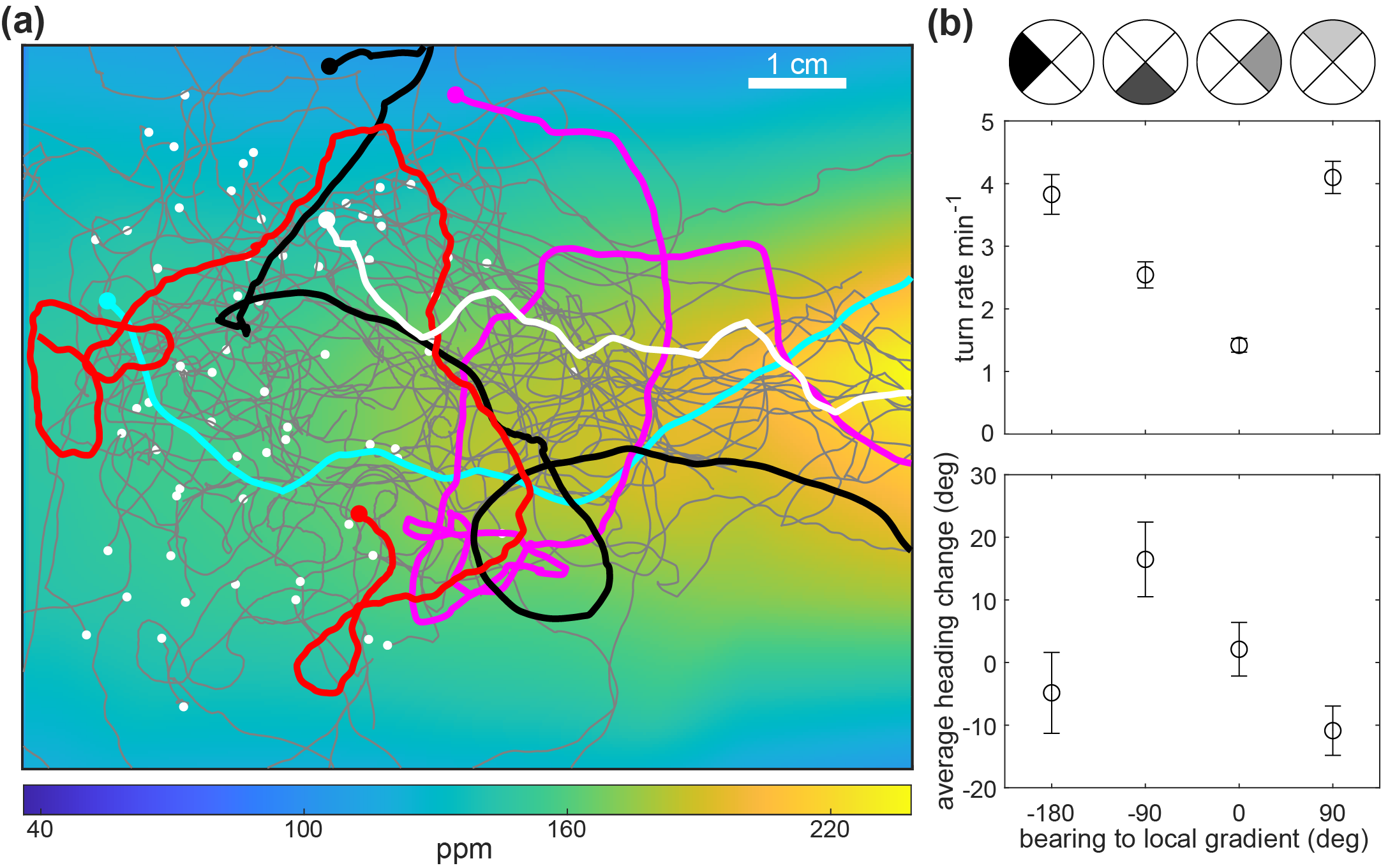}
\caption{
\textbf{\emph{D. melanogaster} larvae chemotaxis in the odor flow chamber.} \textbf{(a)} Trajectories overlaid on the measured butanone odor concentration landscape. Example tracks are highlighted and the initial points are indicated with white dots. \textbf{(b)} Top: turn rate versus the bearing, which is the instantaneous heading relative to the gradient defined by quadrants shown on top. Error bar show counting statistics. Bottom: average heading change versus the bearing prior to all turns (re-orientation with at least one head cast). Error bars show standard error of the mean. Data analyzed from 6 experiments, 59 animals, with 620 turns over 6.8 hours of observation.
\label{fig:fig7}
} %
\end{fullwidth}
\end{figure}

To further evaluate the utility of the flow chamber and gradient calibration for the study of small animal navigation, we investigated how larval \textit{Drosophila} navigate butanone. Although butanone is not as commonly used as a stimulus with \textit{Drosophila} as with \textit{C. elegans}, butanone is known to be attractive to larval flies \citep{dubin1995scutoid, dubin1998involvement} and has been variously reported to be attractive \citep{park2002inactivation} and aversive \citep{israel2022olfactory, lerner2020differential} to adult flies. 
To investigate the larva's navigational strategy in a butanone gradient, we created a "cone" shaped butanone gradient over the agar substrate using the pre-equilibration protocol, as before, and we confirmed the presence and stability of the gradient by continuously measuring the spatial distribution of butanone upstream and downstream of the agar arena. We monitored the orientation and movement of 59 larvae over 6 separate 10 minute experiments ($\sim$ 10 larvae per experiment) with an average observation time of 7 min per larva (\FIG{fig7}). 

Larvae moved towards higher concentration of butanone (\FIG{fig7}a). To analyze the strategy by which they achieved this, we first constructed a coordinate system in which 0 degrees was in the direction of the odor gradient (towards higher concentration) and 180 degrees was directly down-gradient; angles increased counterclockwise when viewed from above.
We found that larvae initiated turns at a higher rate when headed down-gradient ($\pm 180^\circ$ bearing with respect to the local gradient) than up gradient ($0^\circ$) (\FIG{fig7}b). When larvae turned, their reorientations tended to orient up gradient (negative angle changes from $+90^\circ$ bearing with respect to the local gradient, and positive angles changes from $-90^\circ$) (\FIG{fig7}c). Thus \textit{Drosophila} larvae use similar navigational strategies to \textit{C. elegans} to move towards butanone

\section{Discussion}

We present a custom-designed flow chamber and odor sensor array that enables us to measure navigation strategies of worms and fly larvae within the context of a controlled and measured odor environment.
The key features of this odor delivery system are that (1) the odor concentration profile through space is controlled in the flow chamber, (2) the odor sensor array provides a spatial readout to calibrate and measure the profile, and (3) the odor concentration profile is monitored during animal experiments. This last feature, the ability to monitor the spatial profile of odor concentration on the boundary during experiments, sets this method apart from previous approaches. 

The ability to monitor spatial profile during experiments, along with a quantitative understanding of odor-agar interactions, provides confident knowledge of the odor experienced by the animal over time. This in turn allows us to extract tuning curves that describe the animal's behavioral response to the odor it experiences. In the future, such tuning curves may form the basis of investigations into neural mechanisms driving the sensorimotor transformations underlying navigation. 

In contrast to liquid delivery of odor gradients via microfluidic chips \citep{Albrecht2011-fj,Larsch2015-xy}, our method allows worms to crawl freely on an agar surface. This allows our behavior measurements to be directly compared against classical chemotaxis assays \citep{bargmann1993odorant, Louis2008-ju, Pierce-Shimomura1999-nt}. Additionally, the macroscopic odor airflow chamber makes it straightforward to flexibly adjust the the spatial pattern between experiments without the need to redesign the chamber.

Our setup uses low flow rates corresponding to low wind speed velocities (5 mm/s) to avoid anemotaxis.  Larger organisms, including adult flies, navigate towards odor sources by combining odor and wind flow measurements \citep{vergassola_infotaxis_2007, matheson_neural_2022}. Fly larvae exhibit negative anemotaxis at wind speeds 200 to 1000 times higher than those used here \citep{jovanic_neural_2019}, but previous work showed that they do not  exhibit anemotaxis at lower  windspeeds like the ones used here   \citep{Gershow2012-nt}. For example, they do not exhibit anemotaxis  at 12 mm/s which is still higher than the velocities they experience here  \citep{Gershow2012-nt}. Therefore we do not expect \textit{Drosophila} larvae to exhibit anemotaxis under our flow conditions.  \textit{C. elegans} are  not thought  to respond to  airflow. In experiments in aqueous microfluidic chips under flow, \textit{C. elegans}  move towards higher concentrations of attractant and do not respond to the flow of the liquid \citep{Albrecht2011-fj}. In agreement, we do not observe any evidence of \textit{C. elegans} anemotaxis in our chamber in response to  control experiments without odor and with  wind speeds of 5 mm/s \FIGSUPP[fig5]{no_odor_control}.

We focused on the  odor butanone because it is important for a prominent associative learning assay \citep{Torayama2007-qi, kauffman_c._2011}. Butanone is soluble in water, and therefore it interacts strongly with agar. In this work we showed that this   odor-agar interaction makes it challenging to \textit{a priori} infer an odor landscape experienced by the animal when agar is present, but that continuously monitoring the odor profile on the boundary  overcomes this challenge. Other  odors may instead have interactions with other substrates, such as glass, aluminum or plastic, which would  also necessitate the use of our continuous monitoring approach. We show that our system is also compatible  with less water-soluble odors, such as ethanol.

Here we have addressed the problem of creating airborne odor landscapes. The biophysical processes governing odor sensing in small animals such as \textit{C. elegans} are not fully understood. The worm carries a thin layer of moisture around its body as it moves on the agar substrate \citep{Bargmann2006-dy} and it is unclear to what extent the worm pays attention to the concentration of an odorant in the agar below it vs the air above it.
Our reaction-convection-diffusion model suggests that at the quasi-equilibrium conditions used in our experiments the odor concentration in agar is related to the airborne odor concentration directly above it up to a scalar that we predict to be constant across the agar. Although we have not measured this empirically, this suggests that even in the extreme case that the the animal only senses odor molecules in the agar, the odor concentration experienced by the animal in our experiments should differ by no more than a scaling factor compared to our estimates based on the airborne odor concentration.

Knowing the concentration experienced by the animal is not only useful for measuring navigational strategies more precisely than in classical assays, like the droplet chemotaxis assays. It will also be crucial for studying \textit{changes} in navigational strategy, such as those in the context of associative learning \citep{Cho2016-is, Torayama2007-qi}, sensory adaptation \citep{Levy2020-oh, itskovits2018concerted}, and long time scale behavioral states \citep{Calhoun2014-aa, Gomez-Marin2011-ok, klein2017exploratory}. In all those cases, it will be critical to disambiguate slight changes to the odor landscape from gradual changes in the navigational strategies. Continuously monitoring the odor landscape during behavior will remove this ambiguity.

\section{Methods and Materials}

\subsection{Odor flow chamber}

\subsubsection{Flow chamber setup}

The odor chamber (\FIG{fig1}b) was machined from 
aluminum (CAD file in supplementary \nameref{ssec:num1} section). %
The chamber is vacuum sealed with an acrylic lid. The inner arena contains an aluminum insert that can hold the odor sensor array or a square petri dish lid (96x96 mm). The heading in which air flow  can travel above the insert in the arena is 1 cm tall.
The whole setup is mounted on an optical breadboard and enclosed in a black box during imaging.

The airflow system is connected to a pressurized air source, passing through a particulate filter (Wilkerson F08) and a coalescing filter (Wilkerson M03), then regulated by mass flow controllers (MFCs, Aalborg GFC). 
MFCs are controlled via a Labjack D/A board from a computer using custom Labview code. 
We modulate the flow rate bubbling through liquid in enclosed bottles (Duran GL 45). The moisturized or odorized air is then passed into the flow chamber through inlet tubings. The outlets are connected to a copper manifold, then passed to a flow meter to assure that the inlet and outlet flow rate match. An optical flow sensor is fixed on the flow meter to make time stamps for opening and closing of the lid of the flow chamber during animal experiments. A photo-ionization detector (PID, piD-TECH 10.6 eV lamp) is connected to the outlet of airflow, providing calibration for the odor sensor array and detection of air leaks or odor residuals in the system.
Output readings from the PID, MFC,   odor sensors described in the next section, and imaging camera, are all captured on the same computer sharing the same clock. Analog signals from the PID readout and MFC readback are digitized via a Labjack and recorded with the Labview program.

\subsubsection{Odor flow control}
To construct different odor landscapes tubes from the liquid-odor and water reservoirs are connected to the flow chamber in different configurations. 
For a centered "cone-shape" odor landscape the tubing carrying odorized airflow is connected to the middle inlet. For the "biased-cone" landscape, the tubing for odorized air is connected to the inlet 4 cm off-center. For uniform patterns, all are connected to the same source through a manifold. 
For all experiments the background airflow that carries moisturized clean air is set to $\sim$ 400 mL/min, except for \FIG{fig2} where this value was varied, 
The odor reservoir contains either a 11mM or 110mM butanone solution in water with $\sim$ 30 mL/min airflow bubbling through the liquid.

Overall flow rates across the chamber in experiments were always around or less than $\sim$ 400 mL/min to avoid turbulence. We confirmed that this regime had no turbulence by visualizing flow in a prototype chamber using dry ice and dark field illumination. Our empirical observations matched theory: Given that the chamber is 15 cm wide and 1 cm deep, a flow rate up to 1 L/min corresponds to $\sim$ 1.1 cm/s. With kinematic viscosity of air $\sim 0.15$ cm$^2$/s, the Reynolds number is 7.3 times the flow rate in L/min, which is below the turbulence onset (Re=2000). 

\subsection{Odor sensor array}
A spatial array of metal-oxide based gas sensors (Sensorion, SGP30) along with a relative humidity and temperature sensors (ams, ENS210) was used to measure the odor concentration field in the flow chamber. 
Sensors are arranged together into groups of 16 odor sensors and 8 humidity sensors on a custom circuit board (MicroFab, Plano, TX) called an odor sensor bar (OSB). OSB's are in turn  plugged into a second circuit board (OSH Park, Portland, OR) called the odor sensor hub (OSH). OSBs can be added or removed in different arrangements depending on the experiment, for example to make room for agar. Depending on the experiment, up to 112 odor sensors are arranged in a triangular grid such that no sensor directly blocks the flow from its downstream neighbor, accompanied by 56 humidity sensors in a rectangular grid. 

Sensors are read out via the I2C protocol. Each SGP30 sensor has the same I2C address, as does each ENS210 sensor (different from the SGP30); to address multiple sensors of the same type we use an I2C bus multiplexer (NXP, 
PCA9547
). Each OSB contains 2 multiplexers for its 16 sensors. The multiplexers are also addressed over I2C and can have one of 8 addresses (3 address bits). On each board, the two multiplexers share two bits (set by DIP switches); the remaining bit is hardwired to be opposite on the two multiplexers. Thus each OSB can have one of 4 addresses set by DIP switches, and 4 OSBs can be shared on one I2C bus. 

To communicate with the sensors, we used a Teensy 4.0 microcontroller (PJRC, Sherwood, OR) running custom Arduino software. While the Teensy has two I2C busses, we found it more straightforward to use two micro-controllers instead. Both micro-controllers communicated via USB serial to a desktop computer running custom LabView software. Measurements from all sensors are saved to computer disk in real time. Readouts from the humidity sensors are also sent to their neighboring odor sensors in real time for an on-chip humidity compensation algorithm.

\subsubsection*{Heat management}
To avoid generating thermal gradients, the system has been designed to dissipate heat to the optics table. Each metal oxide odor sensor contains a micro hotplate which consumes 86 mW power during readings. To dissipate this heat the aluminum insert inside the flow chamber serves as a heat sink. Odor sensor bars are connected to the insert using heat conductive tape and thermal paste.  The insert and chamber are in turn in direct thermal contact with the optics table. Temperature and humidity is constantly monitored at 8 locations per OSB via the on-board temperature and humidity sensors during experiments to confirm that there is no thermal or moisture gradient created in the environment.

\subsubsection{Measurements and calibration}

We measure from the odor sensors at 1 Hz for both calibration and behavior experiment modes. 
We sample from the PID at up to 13 Hz. We synchronize and time align the measurements from the odor sensor array, MFC read-back, and PID recording with the same computer clock. %

To calibrate the odor sensors to the PID as in \FIGSUPP[fig1]{figSI1-2}, a spatial uniform flow was delivered in a triangle wave or a step pattern.
Time series from each odor sensor and the downstream PID were aligned by time shifting according to the peak location found via cross-correlation. The time shift was confirmed to be reasonable based on first principle estimates form the flow rate. 

After measuring odor sensors' baseline response under clean moisturized air for 5 minutes, an odorized air was delivered. 
To fit calibration curves, the raw sensor readout was fit to the PID measurements with an exponential of form:
\begin{equation}
    \text{PID}(t) = A \exp(B*\text{OS}(t-\tau))
\end{equation}
where $\text{PID}(t)$ voltage is on the left hand side, the scale factor $A$ and sensitivity $B$ are fitted to match the raw sensor reading $\text{OS}(t-\tau)$ that is time shifted by time window $\tau$. This fitted curve maps from raw readings to odor concentration for each sensor. We validate the fitted curve across different recordings. The distribution of the coefficients $A$ and $B$ are relatively uniform across sensors in the middle of the arena. The sensor mapping are also reliable, so using $\pm$std of the fitted curve changes less than $10\%$ of the overall concentration scale of the landscape.

\subsection{Models for odor flow and odor-agar interaction}

We use two models in our work: (1) a convection-diffusion model that captures quasi-steady state odor concentration profile measured without agar used for the fits in \FIG{fig1}f,g and (2) a reaction-convection-diffusion model for odor-agar interaction shown in \FIG{PE_model}. A version of this second model is also used to justify the pre-equilibration protocol, as discussed in the \hyperref[ssec:appendix]{Appendix}.

\subsubsection{Convection-diffusion model for odor flow without agar}

To model odor flow without agar, for example for the fits in \FIG{fig1}f,g, we use a two-dimensional convection-diffusing model:

\begin{equation}
    \frac{\partial C(x,y,t)}{\partial t} = -v\nabla C + D\nabla^2 C \label{eq:convection-diffusion}
\end{equation}
where the  concentration across space and time is $C(x,y,t)$, flow velocity is $v$, and the diffusion coefficient of our odor is $D$. In our chamber, at steady state $(\frac{\partial C}{\partial t} =0)$ we have:
\begin{equation}
    v\frac{\partial C}{\partial x} = D\frac{\partial^2 C}{\partial y^2}
    \label{eq:steady-state}
\end{equation}
because with our configuration flow along the $x$ axis is dominated by convection while flow along the $y$ axis is dominated by diffusion, and therefore $\frac{\partial^2 C}{\partial x^2} \ll \frac{\partial^2 C}{\partial y^2}$.

The fit in \FIG{fig1}f is the solution to equation \ref{eq:steady-state}:
\begin{equation}
\label{eq:2dmodelfit}
    C(x,y) = \frac{C_o}{2}(1-\erf(\frac{x}{2\sqrt{D\frac{x}{v}}})) \exp(-\frac{y^2}{4D\frac{x}{v}}),
\end{equation}
where $\erf$ is the error function and $C_o$ is the odor source concentration measured in air.

In \FIG{fig1}g we show a fit for a one dimensional slice along $y$ at various positions along $x_c$, for the situation in which there is an
odor-source at  $(y=0,x=0)$:

\begin{equation}
    C(y) = \frac{C(x_c,y=0)}{\sqrt{4\pi D \frac{x_c}{v}}} \exp(-\frac{y^2}{4D\frac{x_c}{v}})
\end{equation}
where $\frac{x_c}{v}$ is an analogy of time in non-stationary diffusion process at the cross-section at $x_c$. %

For the fits in \FIG{fig1} the air flow velocity is set to be $v\sim0.5$ cm/s based on the flow rate and geometry of the chamber (15 parallel tubes provide around 450 mL/min of flow into a $\sim$255 mL chamber with $\sim$15 cm$^2$ cross section). The diffusion coefficient $D$ is left as a free parameter and the value that minimizes the mean-squared error between the model and the empirical measurement is used. We chose to leave the diffusion coefficient as a free parameter instead of using butanone's nominal diffusion constant of $D\sim0.08$ cm$^2$/s, because we expect butanone's effective diffusion coefficient to be different in a confined chamber with background flow.   
We note that the fitted profile shown in \FIG{fig1}g,f and the fitted value agrees with what is expected in a stable convection-diffusion process (Peclet number $\sim 80$).

\subsubsection{Reaction-convection-diffusion model for odor-agar interaction}

To justify the pre-equilibration protocol of \FIG{fig4} and to show that measurements of odor concentration along the agar's boundary allows us to infer the concentration on the agar, we propose a reaction-convection-diffusion model. This phenomenological model forms the basis of \FIG{PE_model}. Compared to the convection-diffusion model, we include the "reaction" term to account for odor-agar interactions. 

The model used is a 2D generalization of this non-spatial model:
\begin{equation} \label{eqn_flow+agar}
    \frac{dC}{dt} = -\frac{1}{\tau}(C - C_o) - w \frac{dA}{dt}
\end{equation}

\begin{equation} \label{eqn_agar}
    \frac{dA}{dt} = k_a C (1-\frac{A}{M}) - k_d A 
\end{equation}
where $C$ is a downstream concentration readout after the airflow has surface interaction with the agar gel. The influx odor concentration is $C_o$ and the odor concentration in agar is $A$. Without agar interaction, the flow chamber has its own timescale $\tau$ and the molecular flux into the agar is weighted by a scalar $w$ (so $w=0$ when there's no agar in the chamber). The association and dissociation constants are $k_a$ and $k_d$ and the maximum capacity of odor concentration that can be absorbed is $M$. This model is similar to the description of odorant pulse kinetics shown in \citep{Gorur-Shandilya2019-me}.

In \FIG{PE_model} we use the 2D generalization:
\begin{equation}
    \frac{\partial}{\partial t} C(x,y) = \mathcal{L} C(x,y) - w \frac{\partial}{\partial t} A(x,y)
\end{equation}
\begin{equation}
    \frac{\partial}{\partial t} A(x,y) = k_a C(x,y)(1-\frac{A(x,y)}{M(x,y)}) - k_d A(x,y)
\end{equation}
where $\mathcal{L} =  -v\nabla  + D\nabla^2 $ (\autoref{eq:convection-diffusion}) is a linear operator for the convection-diffusion process and the odor influx is at the boundary $C(x=0,y=0)=C_o$. We perform numerical analysis on the set of 2D equations and permit $A$ to be non-zero only in the region where agar is present. We use a target concentration $C_o$ that is lower than $M$ and $k_a \gg k_d$ to capture odor absorption into agar. In the simulated pre-equilibration protocol we temporarily increase $C_o$ above $M$ then switch back to the target concentration to efficiently reach a steady-state. 

A slightly simplified version of this model forms the basis of the arguments in the \hyperref[ssec:appendix]{Appendix}.

\subsection{Animal handling}

\subsubsection{\emph{C. elegans}}

Wild type \emph{C. elegans} (N2) worms were maintained at 20 C on NGM agar plates with OP50 food patches. Before each chemotaxis experiments, we synchronized batches of worms and conducted measurements on young adults. Worms were rinsed with M9 solution and kept in S. Basal solution for around 30 min, while applying the pre-equilibration protocol to the flow chamber. Experiments were performed on $1.6\%$ agar pads with chemotaxis solution (5 mM phosphate buffer with pH 6.0, 1 mM CaCl$_2$, 1 mM MgSO$_4$) \citep{Bargmann1993-is, Bargmann2006-dy} formed in the lid of a 96x96 mm square dish. 50-100 worms were deposited onto the plate by pippetting down droplets of worms and removing excess solution with kimwipes. The plate was then placed in the odor flow chamber to begin recordings.

\subsubsection{\emph{D. melanogaster}}

Wild type \emph{D. melanogaster} (NM91) were maintained at 25 C incubator with 12 hr light cycle. Around 20 pairs of male and female flies were introduced into a 60 mm embryo-collection cage. A petri dish with apple juice and yeast paste was fixed at the bottom of the cage and replace every 3 hrs for two rounds during the day time. The collected eggs were kept in the petri dish in the same 25 C environment for another 48-60 hours to grow to second instars. We washed down and sorted out the second instar larva from the plate via $30\%$ sucrose in water around 10 min before each behavioral experiments. We used a 96x96 mm lid with $2.5\%$ agar containing $0.75\%$ activated charcoal for larval experiments \citep{Gepner2015-wm, Gershow2012-nt}. Around 10-20 larva were rinsed with water in a mesh and placed onto the agar plate with a paint brush. The same imaging setup and flow chamber configuration as the worm experiments were used for \textit{Drosophila} larva.

\subsection{Imaging and behavioral analysis}

\subsubsection{Image acquisition}

Animals are imaged via a CMOS camera (Baslar, acA4112-30, with Kowa LM16FC lens) suspended above the flow chamber and illuminated by a rectangular arrangement of 850 nm LED lights. The camera acquires $2,500 \times 3,000$ pixel images at 14 fps. A single pixel corresponded to 32 $\mu$m on the agar plate. Labview scripts acquired images during experiments.

\subsubsection{\emph{C. elegans} behavioral analysis}

To increase contrast for worm imaging, a blackout fabric sheet is placed underneath the agar plate. Custom Matlab scripts based on \citep{Liu2018-mv} were used to process acquired images after experiments, as linked in the \nameref{ssec:num2} section. Briefly, the centroid position of worms were found in acquired images via thresholding and binarization. The animal's centerline was found, and its body pose was estimated follwing \citep{Liu2018-mv}, but in this work only the position and velocity was used. The tracking parameters are adjusted for this imaging setup and we extract the centroid position and velocity of worm.

The analysis pipeline focuses on the trajectory of animal navigation in the arena. The trajectories are smoothed in space with a third order polynomial in a 0.5 s time window to remove tracking noise. We only consider tracks that appear in the recording for more than 1 minutes and produce displacement larger than 3 mm across the recordings. Trajectories starting at a location with odor concentration higher than $70 \%$ of the maximum odor concentration in space is removed, since these are likely tracks from animals that have performed chemotaxis already. We calculate the displacement of the center of the worm body in the camera space. The location in pixel space is aligned with the odor landscape constructed with the odor sensor array to compute concentration gradient given a position. To avoid double counting turns when the animal turns slowly, and to mitigate effects of small displacements from tracking noise, we measure the angle change between displacement vectors over 1 s time window and define turns as angle changes larger than 60 degrees. To quantify the curvature of navigation trajectories, we measure the angle between displacement vectors over 1 mm displacement in space. 

\subsubsection{\emph{D. melanogaster} behavioral analysis}
Analysis of fly larvae is performed as previously \citep{Gepner2015-wm, Gershow2012-nt}.

\subsection{Data sharing}
\label{ssec:num1}
Recordings for odor flow control, concentration measurements, and behavioral tracking data are publicly available: \href{https://figshare.com/articles/dataset/Continuous_odor_profile_monitoring_to_study_olfactory_navigation_in_small_animals/21737303}{10.6084/m9.figshare.21737303}

\subsection{Software sharing}
\label{ssec:num2}
\begin{itemize}
    \item Odor sensor array: \href{https://github.com/GershowLab/OdorSensorArray}{https://github.com/GershowLab/OdorSensorArray}
    \item Worm imaging and analysis: \href{https://github.com/Kevin-Sean-Chen/leifer-Behavior-Triggered-Averaging-Tracker-new}{https://github.com/Kevin-Sean-Chen/leifer-Behavior-Triggered-Averaging-Tracker-new}
    \item Larvae imaging: \href{https://github.com/GershowLab/Image-Capture-Software}{https://github.com/GershowLab/Image-Capture-Software}
\end{itemize}

\section{Acknowledgments}
Research reported in this work was supported by the National Institutes of Health  National Institute of Neurological Disorders and Stroke under New Innovator award number DP2-NS116768 to AML and DP2-EB022359 to MHG; the Simons Foundation under award SCGB \#543003 to A.M.L.; by the National Science Foundation, through NSF 1455015 to MHG, an NSF CAREER Award to AML (IOS-1845137),  under Grant No. NSF PHY-1748958 and through the Center for the Physics of Biological Function (PHY-1734030). This work was also supported in part by the Gordon and Betty Moore Foundation Grant No. 2919.02. We thank the Kavli Institute for Theoretical Physics at University of California Santa Barbara for hosting us during the completion of this work.
Strains from this work are being distributed by the CGC, which is funded by the NIH Office of Research Infrastructure Programs (P40 OD010440). We thank the Murthy Lab and Gregor Labs for flies.

\appendix

\begin{appendixbox}
\subsection{Argument for equivalence of with and without agar airborne distributions given equal measurements at the inlet and outlet}
\label{ssec:appendix}

Without agar, the steady-state odor concentration $C_0$ obeys the following dynamics:
\begin{equation}
        \frac{\partial}{\partial t} C_{0}(x,y) = \mathcal{L} C_{0}(x,y),
\end{equation}
and boundary conditions,
\begin{eqnarray}
    \frac{\partial}{\partial y} C_{0}(x,y_{min}) &=& 0 \\
    \frac{\partial}{\partial y} C_{0}(x,y_{max}) &=& 0 \\
     C_{0}(0,y) &=& C_{inlet}(y) \\
     C_{0}(L,y) &=& C_{outlet}(y), \\
\end{eqnarray}
where $\mathcal{L} =  -v\nabla  + D\nabla^2 $ (\autoref{eq:convection-diffusion}) is a linear operator encompassing the convection-diffusion equation. The coordinates $x,y$ are same as \FIG{fig4}, where the flow is along the x-axis and has length $L$.

With agar, assuming the same concentrations on the inlet and the outlet, nearly identical equations hold:
\begin{eqnarray}
    \frac{\partial}{\partial t} C_{1}(x,y) &=& \mathcal{L} C_{1}(x,y) + k_1 C_{agar}(x,y)-k_2C_{1}(x,y) \\
    \frac{\partial}{\partial t} C_{agar}(x,y) &=& k_2C_{1}(x,y) - k_1 C_{agar}(x,y) \\
    \frac{\partial}{\partial y} C_{1}(x,y_{min}) &=& 0 \\
    \frac{\partial}{\partial y} C_{1}(x,y_{max}) &=& 0 \\
     C_{1}(0,y) &=& C_{inlet}(y) \\
     C_{1}(L,y) &=& C_{outlet}(y) 
\end{eqnarray}
where $C_{1}$ is the concentration in the air and $C_{agar}$ the concentration in the agar. The first order interactions are described with rate constants $k_1$ and $k_2$. If the air and agar are in equilibrium at all points ($k_1 C_{agar}(x,y) = k_2C_{1}(x,y)$) then $C_1$ and $C_0$ obey the same equation with the same boundary conditions and are identical.

Assume that $C_1$ and $C_0$ differ by $\Delta C$:
\begin{eqnarray}
    C_1(x,y) &=& C_0(x,y) + \Delta C(x,y) \\
    C_{agar}(x,y) &=& \frac{k_2}{k_1}C_0(x,y) + \epsilon (x,y)
\end{eqnarray}
then
\begin{eqnarray}
    \frac{\partial}{\partial t}\Delta C(x,y) &=& \mathcal{L} \Delta C(x,y) + k_1 \epsilon (x,y)-k_2\Delta C(x,y) \\
    \frac{\partial}{\partial t} \epsilon  (x,y) &=& k_2\Delta C(x,y) - k_1 \epsilon (x,y) \\
    \frac{\partial}{\partial y} \Delta C(x,y_{min}) &=& 0 \\
    \frac{\partial}{\partial y} \Delta C(x,y_{max}) &=& 0 \\
     \Delta C(0,y) &=& 0 \\
    \Delta C(L,y) &=& 0 
\end{eqnarray}

For $\Delta C$ to be non-zero anywhere requires that sources/sinks of excess concentration in the agar $\epsilon (x,y)$ be exactly arranged to exactly cancel out so that $\Delta C$ remains 0 on the boundary, which is unlikely. Further, the mechanism by which a non-zero$\Delta C$ would be created is by transfer of odor from regions of the agar with excess concentration ($\epsilon (x,y) > 0$ to regions with a deficit $\epsilon (x,y) < 0$), so at all points $\frac{\partial}{\partial t} |\epsilon(x,y)| < 0$ and eventually $\epsilon(x,y)$ and hence $\Delta C(x,y)$ must tend to 0 everywhere.

\end{appendixbox}

\end{document}